\documentclass[11pt,preprint]{aastex}
\usepackage{psfig} 

\newcommand{\beq} {\begin{equation}} \newcommand{\eeq}
{\end{equation}} \newcommand{\beqa} {\begin{eqnarray}}
\newcommand{\eeqa} {\end{eqnarray}}

\newcommand{\obs} {{\rm obs}}

 \newcommand{\krho} {{k_\rho}}
\newcommand{\kt} {{k_T}} 
\newcommand{\lt} {\tilde L} 

\newcommand{\nut} {\tilde \nu}
\newcommand{\nch} {{\nu_{\rm ch}}} 
\newcommand{\obss}{{\rm obs}}
\newcommand{\rest}{{\rm rest}}
\newcommand{\rch} {R_{\rm ch}}
\newcommand{\rct} {{\tilde R_c}} 

\newcommand{\ch}  {{\rm ch}}

\newcommand{\rt} {{\tilde r}}
\newcommand{\rmt} {{\tilde r_m}}

\newcommand{\tch} {T_{\rm ch}}

\begin{document}


\shortauthors{Chakrabarti \& McKee}  \righthead{Far-Infrared Spectral Energy
Distributions of Dusty Galaxies}

\title{Far-Infrared Spectral Energy Distributions 
and Photometric Redshifts of Dusty Galaxies}

\author{ Sukanya Chakrabarti\altaffilmark{1,2} \& Christopher
F. McKee\altaffilmark{3}} \altaffiltext{1}{Harvard-Smithsonian Center
for Astrophysics, 60 Garden Street, Cambridge, MA 02138 USA;
schakrabarti@cfa.harvard.edu}

\altaffiltext{2} {National Science Foundation Postdoctoral Fellow}
\altaffiltext{3}{ Departments of Physics and Astronomy, University of California at
Berkeley, Mail Code 3411, Berkeley, CA 94720 USA;
cmckee@astro.berkeley.edu.}

\begin{abstract} We infer the large-scale source 
parameters of dusty galaxies from their observed
spectral energy distributions (SEDs) using the analytic radiative
transfer methodology presented in Chakrabarti \& McKee (2005).  
For local ultra-luminous infrared
galaxies (ULIRGs), we show that the millimeter to far-infrared 
(FIR) SEDs
can be well fit using the standard dust opacity index of 2 when
self-consistent radiative transfer solutions are employed, indicating
that the cold dust in local ULIRGs can be described by a single grain model.
We develop a method for determining photometric redshifts
of ULIRGs and sub-mm galaxies from the millimeter-FIR SED;
the resulting value of $1+z$  
is typically accurate
to about $10\%$.  As such, it is comparable to the accuracy of near-IR
photometric redshifts  and provides a complementary means of deriving
redshifts from far-IR data, such as that from the 
upcoming $\it{Herschel~Space~Observatory}$.  Since our analytic radiative transfer
solution is developed for homogeneous, spherically symmetric,
centrally heated, dusty sources, it is relevant for infrared bright galaxies
that are primarily powered by compact sources of luminosity that are 
embedded in a dusty envelope.  We discuss how deviations from
spherical symmetry may affect the applicability of our solution, and
we contrast our self-consistent analytic solution with standard
approximations to demonstrate the main differences.
\end{abstract} 


\section{Introduction}

The far-infrared (FIR) spectral energy distribution (SED) is a vital
implement in understanding the physical conditions of dusty sources.
Chakrabarti \& McKee (2005, henceforth CM05), 
presented
self-consistent analytic radiative transfer solutions for the spectra
of unresolved, homogeneous, spherically symmetric, centrally heated,
dusty sources.  We showed that from two colors
in the millimeter (mm) and FIR portion of the spectrum
one can approximately infer the mm
to FIR, and that this in turn determines the luminosity to mass ratio, $L/M$, and
surface density, $\Sigma$, which (at low redshift) are distance-independent parameters.
With a distance measurement,
one can further infer the size, mass, and luminosity of the source.  We
extensively compared our analytic solutions against a well-tested
numerical scheme, DUSTY (Ivezic \& Elitzur 1997), to find excellent
agreement with the numerical results.  Here, we apply this methodology
to dusty galaxies to derive their large-scale source parameters.  We 
discuss applications to protostars and radiative transfer methodology of clumpy 
envelopes in a separate forthcoming paper.

From observations of their SEDs, the IRAS all-sky survey characterized ULIRGs as a class of extremely luminous ($L_{8-1000~\micron}>10^{12}L_{\odot}$) galaxies that emit most of their energy in the FIR (Soifer et al. 1984; Aaronson \& Olszewski 1984; Soifer et al. 1987; Sanders \& Mirabel 1996).  These galaxies were then understood to be a new class of objects, quite distinct from those studied by optical surveys as little correlation was found between their optical and infrared luminosities (Sanders \& Mirabel 1996).  That the demographics of these galaxies is not simply an extrapolation of normal galaxies can be seen from the fact that the luminosity function on the bright end ($L_{\rm IR} \ga 1 \times 10^{11} L_{\odot}$) is significantly in excess of the Schechter function (Rieke \& Lebofsky 1986).  Theoretical models have suggested that heavily starbursting systems like ULIRGs can be produced via mergers of roughly equal mass galaxies (Toomre \& Toomre 1972; Mihos \& Hernquist 1996), and recent observations corroborate the idea that local ULIRGs are products of major mergers (Dasyra et al. 2006).  While the limited sensitivity of IRAS did not allow for a characterization of the redshift evolution of this dusty, luminous population of galaxies, observations by the $\it{Spitzer~Space~Telescope}$ indicate that there is a strong evolution in this population (on the basis of near and mid-IR observations converted to total IR luminosities using observational templates) out to $z \sim 1$ (e.g. Le Floch et al. 2005), with the contribution from ULIRGs to the comoving IR energy density increasing by an order of magnitude from local systems to $z \sim 1$.

Understanding the FIR SEDs of dusty galaxies has
renewed importance today.  Submillimeter galaxies (SMGs; $F_{850~\micron} \ga 1~\rm mJy$) (Ivison et al. 2000; Blain et al. 1998) are luminous ($L \ga 10^{12} L_{\odot}$), dusty galaxies 
at moderate redshifts (the median redshift  in the Chapman et al. 2003, 2005
samples is $z \sim 2$).  
They are faint at optical wavelengths and
were discovered in the first deep extragalactic surveys in the sub-mm
wavebands (the SCUBA Cluster Lens Survey; Smail et al 1997, 2002).
SMGs produce a significant fraction of the energy output of the high 
redshift early universe, and hence represent a cosmologically significant population
(Smail et al 1997, Blain et al 2002, Blain et al 1999).  Chapman et al. (2005) find 
that SMGs and Lyman break galaxies contribute equally to the star formation density 
at $z \sim 2-3$ and that,
when extrapolated to lower fluxes,
 SMGs may be the dominant site of massive star formation at this epoch.  
Upcoming instruments, such as the $\it{Herschel~Space~Observatory}$ and SCUBA-2, will be able to perform routine observations of SMGs at rest-frame FIR wavelengths, which is critical for observationally determining the bolometric luminosities of this high redshift, cosmologically significant population.  In
contrast to local ULIRGs, the suggested formation mechanisms and evolutionary scenarios for SMGs remain varied in nature, ranging from primeval, heavily accreting galaxies undergoing a starburst (Rowan-Robinson 2000; Efstathiou \& Rowan-Robinson 2003) to products of gas-rich major mergers undergoing intense feedback (Chakrabarti et al. 2007b), with recent observations favoring the latter scenario (Nesvadba et al. 2007; Bouche et al. 2007).  In particular, Bouche et al. (2007) suggest
that dissipative major mergers may have produced the SMG population on the basis of their
finding that the SMG population has lower angular momenta and higher matter densities
compared to the UV/optically selected population.

A self-consistent analytic 
method of inferring source parameters from the observed SEDs may be an useful alternative to SED templates in analyzing upcoming FIR data sets.
We give the general relations for observed quantities in terms of the redshift, and graphically depict the variation of these quantities
with redshift, which is significant even at $z\sim 1$.  
We show that this 
implies that one can estimate 
the value of $1+z$ for ULIRGs and
SMGs from the mm and FIR SED with an accuracy $\sim 10\%$ (this is comparable to 
typical accuracies of photometric redshift codes, which estimate $z$ to typical accuracies
of $\sim 10\%$, e.g. the IMPZ code of Babbedge et al. 2004, or the widely used HYPERZ code of Bolzonella et al. 2000),  given that 
our assumptions 
are satisfied: (1) the mm-FIR spectrum is due to reprocessing of emission from
a central, dust-enshrouded source; (2) the dust can be approximated as being
homogeneous and spherically distributed, with a density that varies
as a power of the radius; (3) the source is sufficiently opaque that emission
from the dust destruction front is negligible; and (4)
the luminosity-to-mass ratio, $L/M$, of high-redshift ULIRGs and SMGs is similar to
that of low-redshift ULIRGs (this last assumption is verified for the small set of
SMGs for which adequate data currently exist).
Chapman et al. (2005) point out that the dust temperatures inferred for SMGs are
significantly lower than those of local ULIRGs and conclude that FIR photometric
redshifts have an uncertainty $\Delta z \simeq 1$; as we shall see, our method is 
significantly more accurate.  We primarily focus our
discussion of SMGs on sources observed recently by Kovacs et al. (2006)
using the $350~\micron$ band of SHARC-2, which is currently the most direct
observational probe of the rest-frame FIR of high redshift SMGs.

The organization of the paper is as follows: in \S 2, we review the
basics of the analytic radiative transfer methodology presented in CM05
and collect the main expressions in Appendix A; in \S 2.1, we explain
the general procedure of applying our results, and contrast our
solution with standard approximations in Appendix B. \S 3 is devoted to a treatment
of ULIRGs, where we infer the large scale parameters of a dozen local
ULIRGs by fitting to the FIR SED.  In \S 4 we present SED fits for a sample 
of SMGs.   
In \S 5 we present the principal result of this paper, a method of
inferring redshifts from FIR SEDs, and we 
demonstrate its applicability
both with a simulated test case and with data of SMGs.  We conclude
in \S 6. 
In Appendix B
we discuss standard approximations, such as the Hildebrand (1983)
prescription for the mass in terms of the mm flux, and modified
blackbody single temperature models in fitting the SEDs of ULIRGs and
protostars (Yun \& Carrilli 2002, henceforth YC02).  For purposes of
illustration, we graphically contrast our solution and the standard
approximations against the numerical results from DUSTY over the
astrophysical parameter space.

\section{Analytic SEDs of Dusty Sources}

In CM05, we formulated an analytic solution for the FIR SEDs of
spherically symmetric, homogeneous dusty sources with a central source
of radiation.  We considered envelopes that emit most of their
radiation at wavelengths longer than 30 $\mu\rm m$, and are sufficiently opaque
that emission from the dust destruction front does not
significantly influence the FIR SED.  A corollary to this
assumption is that the emergent spectrum is also approximately
independent of the temperature of the source of radiation.  We did not
consider the effects of scattering on the SED, since the scattering
efficiency is much smaller than the absorption efficiency at mm and
FIR wavelengths.  Here, we give a summary of the solution presented in CM05
and collect our expressions for the analytic SED in Appendix A.

We consider power law density variations within the envelope,
$\rho(r)\propto r^{-k_{\rho}}$ and
 adopt the dust opacity, $\kappa_{\nu}$, from Weingartner \& Draine (2001,
 hereafter WD01), with a
 normalization appropriate for water ice mantles (see eq. \ref{eq:kappa})
For a given density variation in the envelope and dust opacity curve, the
emergent spectrum depends on three quantities - the mass of the envelope:
$M_{\rm dust}=M/Z_{\rm dust}$, the luminosity, and the radius of the
envelope: $R$.  The shape of the SED cannot depend on the
distance to the source, and can be specified by two
distance-independent parameters, the luminosity to mass ratio, $L/M$,
and the surface density, $\Sigma\equiv M/\pi R^{2}$. 

We defined characteristic parameters, $\rch$ and $\tch$ that
are analogous to the Rosseland photosphere and photospheric temperature
respectively, such that 
\beq L\equiv 4\pi\lt \rch^{2}\sigma \tch^{4}\;,
\label{eq:L} \eeq 
where $\sigma=5.67\times 10^{-5} \rm
~erg~cm^{-2}~s^{-1}~K^{-4}$ is the Stefan-Boltzmann constant.  $\lt$
is a number of order unity that allows for better agreement
with the numerical solutions, particularly for extended atmospheres,
which have $\lt \ga 1$, reflecting the effective increase in emitting
area.

The characteristic parameters are determined by requiring that the
characteristic optical depth at a frequency $\nu_{\rm ch}\equiv
kT_{\rm ch}/h$ equal unity:
\begin{equation} \tau_{\rm ch}=\kappa_{\nch}
\int_{1}^{{R_c}\rightarrow\infty}\rho(\tilde{r})d\tilde{r}=
\frac{\kappa_{\nch} \rho(\rch)R_{\rm ch}}{k_{\rho}-1}=1\; ,
\label{eq:tauch}
\end{equation} 
where $\kappa_{\nch}$ is the opacity per unit mass at $\nch$ and
$\tilde r\equiv r/\rch$. Note that this characteristic optical depth
is for an effectively infinite shell; it does not take into account the
edge of the core at $R_c$.  
We assume that $\nu_{\rm ch}$ is
within the frequency range where the opacity is approximately a power
law: 
\beq
\kappa_{\nu}=\kappa_{\nu_{0}}(\nu/\nu_{0})^{\beta}~~~~~(30~\mu{\rm m
\la \lambda \la 1~mm}).
\label{eq:kappa}
\eeq
For $\nu_0=3$~THz, corresponding to $\lambda_0=100~\micron$,
we adopt an opacity per unit mass of gas of
\beq
\kappa_{100\,\micron}=0.54\delta ~~~\mbox{cm$^2$ g$^{-1}$}.
\eeq
For $\delta=1$, this is twice the value given by 
WD01 model for dust in the diffuse interstellar medium
since  we assume that grains in star-forming regions have ice mantles that
double the FIR opacity.
The WD01 opacity is based on a gas-to-dust mass
ratio of $M/M_d=105.1$; since the ice mantles most likely have
a different opacity per unit mass than the WD01 grains, we
do not attempt to infer the dust mass in the sources.
Deviations from solar metallicity, or from the
assumed dust model, can be taken into account by choosing
a different value for $\delta$.

Solving equations (\ref{eq:L}) and (\ref{eq:tauch}) gives the
relations between the source parameters, $L/M$ and $\Sigma$, and the
SED variables, $\rct$ and $\tch$ that govern the shape of the SED:
\begin{equation} 
\rct\equiv\frac{R_{\rm c}}{R_{\rm ch}}=
\left\{\frac{(L/M)\Sigma^{(4+\beta)/\beta}}{4\sigma \tilde{L}}
\left[\frac{(3-k_\rho)\kappa_{\nu_0}}{4(k_{\rho}-1)T_0^\beta}\right]
^{4/\beta}\right\}^{-\frac{\scriptstyle \beta} {\scriptstyle
2\beta+4(k_{\rho}-1)}}\; ,
\label{eq:rct}
\end{equation} 
and
\begin{equation} 
T_{\rm ch}=\left\{\frac{L/M}{4\sigma \tilde{L}
\Sigma^{\frac{3-k_{\rho}}{k_{\rho}-1}}}
\left[\frac{4(k_{\rho}-1)T_{0}^{\beta}}{(3-k_{\rho})\kappa_{\nu_{0}}}\right]
^{\frac{2}{k_{\rho}-1}} \right\} ^{\frac{\scriptstyle
k_{\rho}-1}{\scriptstyle 2\beta+4(k_{\rho}-1)}}\; ,
\label{eq:tch}
\end{equation} 
where $kT_0\equiv h\nu_0$.
We find that
\begin{equation} \tilde{L}=1.6\rct^{0.1} \;
\label{eq:ltil_val}
\end{equation} 
is accurate to within $\sim$ 10 \% for $1\la \krho\la 2$;
note that this value is about twice the value given in CM05.
Equations (\ref{eq:rct}) and (\ref{eq:tch})
allow us to $\it{analytically}$ solve for the distance-independent
source parameters once the SED variables are determined from two 
colors (i.e., flux ratios).
We depict in Figure 1 the $L/M$ vs $\Sigma$ plane overlaid
with lines of constant $\rct$ and $\tch$ for the density
profile, $k_{\rho}=3/2$ and adopted dust model, WD01 coated with ice mantles.

We model the emergent spectrum by assuming that 
the emission in each
frequency channel comes from a shell of thickness $\Delta r_{m}(\nu)$
centered at a radius $r_{m}(\nu)$, with a source function
$(2h\nu_{ch}^{3}/c^{2})\exp\left[{-h\nu/kT(\rmt)}\right]$ located at
an optical depth $\tau_\nu(\tilde{r}_{m})$:
\begin{equation} L_{\nu}=4\pi
\rch^{2}4\pi\left(\frac{2h\nu_{ch}^{3}}{c^{2}}\right)
\tilde{\kappa}_{\nu}\tilde{\nu}^{3}(k_{\rho}-1)\tilde{r}_{m}^{2-k_{\rho}}
\Delta\tilde{r}_{m}\exp\left[-\frac{h\nu}{kT(\tilde{r}_{m})}-
\tau_\nu(\tilde{r}_{m})\right]\; ,
\label{eq:shell}
\end{equation} where the optical depth $\tau_\nu$ from $r$ to the
surface of the cloud is
\begin{equation}
\tau_{\nu}=\tilde{\kappa}_{\nu}\left(\tilde{r}^{-k_{\rho}+1}-
\rct^{-k_{\rho}+1}\right) \; .
\label{eq:tau}
\end{equation}

The FIR emission can be represented with good accuracy with the
adoption of a power law for the temperature profile, 
\beq
T=
\tch\rt^{-\kt}\; .
\eeq 
The
slope of the temperature profile is determined by imposing the
self-consistency condition that the input luminosity exactly equal the
emergent luminosity. As expected, the slope of this effective
temperature profile is independent of the optical depth in the limit
of low optical depths, and becomes progressively a steeper function of
the optical depth for very opaque envelopes.  
For envelopes that emit most of their flux at wavelengths longer than
$30\mu\rm m$, we showed that $k_{T}$ is a function of $\rct$ only.
The functional forms of $k_{T}(\rct)$ and $\tilde{r}_{m}$ are given in
Appendix A.

We found that spectra are characterized by three frequency regimes,
which we denoted as low, intermediate and high.  Low and
intermediate frequencies are optically thin.  Low frequencies are in
the Rayleigh-Jeans portion of the spectrum [$h\nu\ll kT(\rct)$]
throughout the envelope.  The low-frequency emission comes
predominantly from the outer parts of the shell and is proportional
to the mass.  Intermediate frequencies are in the Wien part of the
spectrum in the outer envelope, but not near the photosphere.  
In the intermediate frequency regime, higher frequencies
originate from deeper in
the envelope, where it is hotter.
High frequency photons emanate from a
location in the shell that is due to a tug-of-war between the
temperature gradient, which favors emission from small radii,
and the intervening optical depth, which favors emission from large radii.  
The forms of the characteristic emission radius, $\rmt$
(termed the ``contribution function'' in CM05) and the thickness, $\Delta
\rmt$, are given in Appendix A.  
We illustrate these frequency regimes in 
Figure 2a with the example of a high-mass protostar from Paper I 
($L/M \sim 400
L_{\odot}/M_{\odot}$ and $\Sigma \sim 1~ \rm g\;cm^{-2}$); both the
analytic and numerical versions of the SED are shown.
Figure 2b
shows the characteristic emission radius (contribution function) as
a function of frequency, and
Figure 2c is a plot of the opacity curve, WD01's $R_{V}=5.5$
(this curve does not include the effect of ice mantles). 
One should read these three plots left to right, i.e., follow the marked
regions in the SED plot in Figure 2a and correlate them with the
marked regions in the contribution function in Figure 2b.  The
spectral features in the contribution function in Figure 2b correlate
with the spectral features in the opacity curve as depicted in Figure
2c.  For example, the 10 $\mu$m ($3\times 10^{13}$ Hz) increase in the
opacity translates to a corresponding increase in $\tilde{r}_{m}$, as
the $\tau=1$ surface at this frequency is driven outwards, while the 5
$\mu$m ($6\times 10^{13}$ Hz) decrease in the opacity causes
$\tilde{r}_{m}$ to move inwards.

\subsection{Inference of Source Parameters}

We solve equations (\ref{eq:rct}) and (\ref{eq:tch}), inserting the
relation for $\lt$ from equation (\ref{eq:ltil_val}), to give the
source parameters in terms of SED variables, $\tch$ and $\rct$:
\begin{equation} 
\frac{L}{M}=1.6\left(\frac{3-\krho}{k_{\rho}-1}\right)\kappa_{\nu_0}\left(\frac{\sigma T_{\rm ch}^{\scriptstyle 4+\beta}}{T_0^\beta}\right)\rct^{\scriptstyle k_{\rho}-2.9}
\; ,
\label{eq:LM}
\end{equation}
\begin{equation}
\Sigma=\frac{4(k_{\rho}-1)}{(3-k_{\rho})}\frac{1}{\kappa_{\nu_{0}}}\left({\frac{T_{0}}{T_{\rm
ch}}}\right)^{\beta}\tilde{R_{c}}^{\scriptstyle -(k_{\rho}-1)} \; .
\label{eq:Sigma}
\end{equation} 

Since $\kappa_{\nu_0}$ is proportional to the dust-to-gas parameter $\delta$,
it follows that we actually infer values for $L/(M\delta)$ and $\Sigma\delta$
from the SED---i.e., the SED is determined by the mass of dust, not gas, in the source.
We shall refer to these scaling relations throughout
the course of the paper.
Our method for inferring the source parameters depends
on whether the redshift is known or not. In either case, 
we require at least three photometric data points, with at
least one in or near each of the three frequency regimes.
For example, if there are no data at low frequencies or in
the transition region between low and intermediate frequencies, then we
can determine only a lower limit on $\rct$; if there are no data at
high frequencies or in the transition region between intermediate and
high frequencies, then we can determine only a lower limit on $\tch$.
If additional data are available we infer the best-fit
value for the density profile, $\krho$, as well. We generally find
$\krho=2$ for ULIRGs as we show in \S 3, so we adopt this value if additional data are not available.

	First consider sources with known 
spectroscopic redshifts. For such sources, we transfer the observed data to the rest frame:
\beq
\nu_{\rm rest}=\nu_{\rm obs}(1+z),~~~~~
L_{\nu,\rm rest} = \frac{4 \pi D_{L}^2 F_{\nu,\rm obs}}{1+z}\; .
\label{eq:rest}
\eeq
Using the rest-frame values of $L_\nu$, we solve for $\rct$ and $\tch$ and determine 
the shape of the SED; this allows us to determine the total luminosity, $L$.
Equations (\ref{eq:LM}) and (\ref{eq:Sigma}) allow us to infer $L/M$ and $\Sigma$,
which in turn give us $M$, $R_c=(M/\pi\Sigma)^{1/2}$, and $\rch=R_c/\rct$.

	For sources with unknown redshifts, one can fit to the observed-frame fluxes and solve
for $\rct$ and $\tch$, and then obtain  $(L/M)_\obss$ and $\Sigma_\obss$ via
equations (\ref{eq:LM}) and (\ref{eq:Sigma}).  
As we shall see in \S 4 below, the observed-frame values of $L/M$ and $\Sigma$
are quite different from the intrinsic, rest-frame values, and it is possible to
estimate the redshift from $(L/M)_\obss$. With an approximate redshift, one
can then infer approximate values for the source parameters, as described above.

	We can also approximately obtain the
angular size of the photosphere of the sources, even if the redshifts are unknown.
The angular size of the photosphere is about $\theta_{\rm
ch}\equiv \rch/D_{A}$, where $D_{A}$ is the angular
diameter distance.  We use equation (\ref{eq:L}) and the relation between the angular diameter
distance and the luminosity distance: $D_{L}=(1+z)^{2}D_{A}$, to write the
angular size in terms of the total flux, $F\equiv L/4\pi D_{\rm L}^{2}$:
\beq
\theta_{\rm ch}=(1+z)^{2}\left(\frac{F}{\tilde{L} \sigma \tch^{4}}\right)^{1/2}=\left(\frac{F}{\tilde{L} \sigma T_{\rm ch,obs}^{4}}\right)^{1/2}\; .
\eeq
Thus, we can infer the angular size of the source even
if it is unresolved, i.e., we predict the size of the source from
observed quantities
without knowing the redshift (except for the very weak dependence on
$\tilde{L}$).  

Many of the sources we consider in this paper may have high-frequency 
fluxes ($\lambda \la 30
~\micron$) fluxes that 
are affected by an inhomogeneous dust envelope, multiple sources of
luminosity, and/or an
accretion disk. All of these considerations lie outside the scope of
the methodology developed in CM05, and to avoid them we perform fits
to the mm - 30 $\micron$ data, which are 
unlikely to have been affected.
If the resulting model for the SED fits the high-frequency fluxes as
well, then that
is a good indication that the high frequency emission from the source
can also be well-described by our simple theoretical construct.

If the envelope is optically thick at FIR wavelengths and the slope of the temperature profile,
$k_T$, is a
strong function of $\rct$, one cannot analytically isolate the
unknowns $\tilde{R_{c}}$ and $\tch$ to express the observed color
ratios in terms of the SED parameters.  In this case, we
have a set of coupled algebraic equations that must be numerically
solved, including the uncertainties on the data points, using a
least-squares routine.   
As discussed in CM05, the accuracy for our analytic
SED over the astrophysical parameter space, for 
$1\mbox{ mm}>\lambda \ga 30
~\mu\rm m$, is generally $\la 30\%$.  Our accuracy in the FIR
($300-60 ~\mu\rm m$) is typically $\sim 10\%$.  If the reported errors
on the data points are less than 30\% at millimeter wavelengths and
less than 10\% elsewhere in the spectrum, 
we artificially increase them 
to these values in order to account for the intrinsic error in the analytic solution.
We report the reduced chi-squared values for both the reported errors
and the rescaled errors, if these two values differ dramatically.
Rescaling the observed errors to account for the intrinsic error in
the analytic solution is analogous to taking an upper bound on the
uncertainty.  
Since the errors in the analytic SED are correlated,
the effective uncertainty due to the intrinsic error in the
analytic solution and the reported errors for the observed data is
less than or equal to the standard deviation of the errors.

As noted in CM05, there are two features of our solution that have not
been previously emphasized in the literature.  We describe them here in 
terms of rest-frame quantities.  Firstly, extended
envelopes (large $\rct$) have a three-component spectrum such that the
intermediate frequency component separates cleanly from the low
frequency component.  The frequency at which this break occurs
is termed the break frequency. 
For large $\rct$, it is given by (see eq. 21 in CM05;
note that there is a typographical error, and that the sign of the exponent of
$\rct$ should be negative), 
\beq 
\tilde{\nu}_{\rm
break}=\left[2.5(2.6-k_{\rho})\Gamma\zeta\right]^{\frac{\scriptstyle 0.4}{\scriptstyle 2.6-k_{\rho}}}\rct^{-0.4}\; ,
\label{eq:nubreak1} 
\eeq 
where the argument of the Gamma and Zeta
functions, $(3-k_{\rho})/k_{T}$, has been suppressed for clarity, and
where we have set $k_T=0.4$, which is appropriate for large $\rct$.
We shall find below that the density profile typically corresponds to $\krho=2$;
in that case, $\tilde\nu_{\rm break}=1.93\rct^{-0.4}$. The break frequency is
comparable to the frequency that corresponds to the temperature at the outer edge of the source,
$\nu_c\equiv kT_c/h$, since $\tilde\nu_c=T_c/\tch=\rct^{-k_T}$; for the typical case,
it is $\nu_{\rm break}=1.93\nu_c$.
The frequency dependence in the intermediate frequency regime 
($\nu_{\rm break}<\nu\la\nch$)
for large $\rct$ is given
by: 
\beq 
L_{\nu}\propto \nu^{3+\beta-2.5(3-k_{\rho})}\; .
\eeq

Secondly, we presented the ratio of the peak frequency of the SED
(expressed as $F_\nu$)
in terms of the characteristic frequency.  For compact envelopes (low
$\rct$), our results are similar to the blackbody limit, with
$\nu_{\rm peak}\sim 3\nu_{\rm ch}$, while for extended envelopes, the
peak frequency tends to the characteristic frequency.  This variation
can be approximated by the following form: 
\beq 
\frac{\nu_{\rm
peak}}{\nu_{\rm ch}} \simeq 0.82k_{\rho}+\frac{5.4-1.8k_{\rho}}{
\rct^{0.56k_{\rho}-0.22}}\; ,
\label{eq:nupk3} 
\eeq
for $1\la \krho\la 2$ to within $\sim$ 20 \% accuracy, for $5000 \ga \rct \ga 10$.

\section{SEDs \& Inferred Parameters for ULIRGs}

We present SED fits and source parameters for ten well-known local ULIRGs.
Our methodology is most applicable to ULIRGs predominantly powered by a source of luminosity that can be approximated as being compact and
nearly enshrouded by dust.  Soifer et al. (1999; 2000) found that a
large fraction of the mid-infrared emission in a sample of the closest
ULIRGs stems from very compact (few hundred parsec) systems, rather
than from
extended (kiloparsec), weakly obscured starbursts.  

High-resolution imaging has revealed many of the complex
geometries of merging galaxies (Soifer et al 2000, Goldader et al
2002, Downes \& Solomon 1998, Bushouse et al 2002, Scoville et al
1998, Soifer et al 1999, Scoville et al 2000, Surace et al 2000,
Surace \& Sanders 1999) - in our simple treatment here, we cannot
consider these intricate features.  Our goal here is to understand the
large-scale characteristics of these systems from their FIR
SEDs, by approximating them as spherical dust envelopes surrounding a
compact central source of luminosity.  
Recent numerical work by Siebenmorgen \& Krugel (2007) on modeling
the SEDs of ULIRGs suggests that deviations from spherical symmetry
may not significantly affect the FIR SEDs of these systems.  Chakrabarti et.
al (2007a) solved for the SEDs of ULIRGs using a self-consistent three-dimensional
radiative transfer code that takes the gas and stellar densities
as input from smoothed particle hydrodynamics simulations, 
and found that large-scale trends in the 
FIR SEDs of ULIRGs can by described in terms of these two basic parameters, 
$L/M$ and $\Sigma$, as discussed originally by CM05.

The dust-to-gas ratios in ULIRGs are not entirely certain.  
The work by Dunne \& Eales (2001) found that using two-temperature
fits to SEDs leads to dust-to-gas ratios that are closer to Milky Way
values than previous work, based on single-temperature fits, had found.
Farrah et al. (2005) find (slightly) super-solar metallicities
from Space Telescope Imaging Spectrograph (STIS) observations of
star-forming knots of dense gas in the nuclear regions of ULIRGs.
There are no definitive claims of metallicity gradients in ULIRGs, though
one may expect the nuclear regions to be enriched relative to the outer
regions; for example,  Sodroski et al. (1997) found that the
gradient of the dust-to-gas mass ratio is
comparable to the metallicity gradient in the Milky Way.  Such gradients are also
likely to exist in ULIRGs. 
We quote our results using the local interstellar dust-to-gas ratio
(1/105.1, corresponding to $\delta=1$), but tabulate $M\delta$ so that the results can
be readily scaled to different dust-to-gas ratios.

As noted previously, we have used the WD01 dust model, which has a long
wavelength dust opacity slope, $\beta=2$, in performing the fits.
Our results show that this fits
the long-wavelength points reasonably well 
(see Figures 3-5).  
On the other hand, using single-temperature modified
blackbody models (e.g. YC02) generally requires $\beta< 2$ to fit
the long-wavelength slopes of ULIRGs.
Our results for a continuous temperature distribution are consistent with those of
Dunne \& Eales (2001), who showed that two-temperature blackbodies
are enough to fit the spectra with $\beta=2$. 
Note that the observed spectrum does not have a slope of 2
in the intermediate frequency regime ($\nu_{\rm break}\la\nu\la\nu_{\rm pk}$;
see eqs. \ref{eq:nupk3} and \ref{eq:nubreak1}).

We discuss some of the ULIRGs in Table 1, beginning with Arp 220, which
at a distance of $77 \rm ~Mpc$ 
is the nearest ULIRG and 
has received the most scrutiny. The
nuclear region of Arp 220 has been resolved into a double nucleus
(Graham et al 1990) with associated gaseous disks (Downes \& Solomon
1998; henceforth DS98) and interpreted to be the result of a merger.  The
FIR SED of Arp 220 has $F_{100\mu\rm m}\sim F_{60\mu\rm m}$, 
$\tch \sim 125\rm~K$, and is broad, with $\rct \sim 370$, $\rch=30\rm~pc$ for $\krho=2$. These SED variables give $\Sigma \sim 0.03\rm~g~cm^{-2}$.  
We estimate a size $R_c\simeq 11
\rm ~kpc$ and $\rch=30\rm~pc$ for a density profile, $k_{\rho}=2$, at a
confidence level of 86 \%.  
The large outer radius is needed to fit the mm data.
If we allow the density profile parameter to vary, we find that
it is constrained to a narrow range of values, $1.7 \la \krho \la 2.1$,  with 
the chi-squared per degree of freedom increased by at least
one outside these bounds.  

From analysis of images of Arp 220 at wavelengths
between $3-25\rm~\micron$, Soifer et al. (1999) noted that fluctuations in
seeing could increase the apparent size of the source at
these wavelengths, and concluded that the reported FWHM at
$24~\micron$ of $0.73''~(270~\rm pc)$ (the largest measured diameter
among the various wavelengths) could in fact be as small as
$0.25''~(90~\rm pc)$.
We use the analytic methodology 
of CM05 to compute the characteristic emission radius at $24~\micron$ to find that most of the
$24~\micron$ flux is coming from $r\la r_m(24\;\mu{\rm m})=185~\rm pc$, 
which is about
seven times the characteristic radius.  
This should be understood as an approximate
estimate of the characteristic emission radius at this wavelength, as a clumpy geometry may 
begin to influence the emergent spectrum somewhat for $\lambda \la 30~\micron$;
nonetheless, it is consistent with the observations.
Early studies of the CO and millimeter dust continuum
emission in Arp 220 (e.g. Scoville et al 1991) noted the existence of
an extended component, with a size of $7''\times 15''~(2.6~\rm kpc
\times 5.6~\rm kpc)$, which is in rough agreement with the
overall source
size we have derived from the SED.

Dunne \& Eales (2001) fit the FIR SED of Arp 220 with a
two-temperature blackbody and $\beta=2$, noting that the masses they
infer with the two-temperature fits are a factor of $\sim
2$ larger than those inferred with single-temperature fits.
Our estimate of the outer core temperature,
$T_{\rm c}=13 \rm ~K$, is close to their cold-dust temperature of $18
\rm ~K$ for Arp 220.  In contrast, YC02 fit the FIR SED of Arp 220
with a single-temperature modified blackbody, finding a best-fit
$\beta=1.1$.  Our estimate of the mass ($ 4\times 10^{10}\;
M_{\odot}$) is in close agreement with the estimate by Dunne \&
Eales (2001). Appendix B gives a 
general explanation for the larger gas masses obtained by Dunne \& Eales (2001) 
and ourselves when using two temperatures or a continuous range of
temperatures in fitting the SEDs of extended (large $\rct$) 
envelopes, in contrast to using a single-temperature blackbody model.  
DS98 estimated gas masses using a model of subthermally excited CO to fit
CO interferometric observations, and derived dynamical masses
from their measured line widths and measured CO radii from maps
of the CO emission in local ULIRGs.   DS98's measurements provide
an independent confirmation that the gas density profile varies as
$r^{-2}$.  They estimate a gas mass interior to 1.36 kpc 
of $5.2 \times 10^{9}~M_{\odot}$, while we estimate a gas mass
of $4 \times 10^{10}\delta^{-1}
~M_{\odot}$ over a radius of 10.3 kpc.  
Since $M \propto r$ for $\rho \propto r^{-2}$,  we infer
a gas mass $M(<1.36\;\mbox{kpc})= 5.3 \times 10^{9} \delta^{-1} M_\odot$,
which is in excellent agreement with their result for $\delta=1$.
These estimates of the gas mass are 
less than the cited dynamical
mass, which in the inner kiloparsec region is dominated by
the stars.

	It is clear however, that we have significantly underestimated the
flux for $\lambda \la 30 ~\mu\rm m$ .  The mid-IR emission is strongly
temperature and opacity dependent and would be significantly
influenced by the clumpiness of the dust envelope, effects of geometry
(the presence of a disk-like structure), and contributions from
distributed sources of luminosity, none of which we have accounted
for.  The contribution from weakly obscured starbursts appears to be
ruled out from previous studies (e.g., Soifer et al. 1999; Soifer et al. 2000).
We consider the effects of inhomogeneities of the dust envelope and their effects on the
high-frequency part of the spectrum in a subsequent paper.

UGC 5101, at a redshift of $z=0.04$, is thought to contain a buried
AGN, based on X-ray observations (Imanishi et al 2003), analysis of
PAH and ice-absorption features (Imanishi \& Maloney 2003, Imanishi et
al 2003), and due to the apparent compactness seen in high resolution
imaging (Scoville et al 2000, Soifer et al 2000).  
$\it{Spitzer}$ IRS observations of fine structure lines, in particular
the high-ionization potential line NeV (the production of which requires 
energies greater than can be produced by OB stars) further confirm the presence of a buried AGN in
this source (Armus et al 2004).  UGC 5101 belongs to the new class of
XBONGs (X-ray Bright Optically Normal Galaxies) discovered by recent
X-ray observations (Maiolino et al 2003, Comastri et al 2002); these
surveys uncovered a group of optically elusive AGN that do not show
a Seyfert-like spectrum in the optical, but do have a hard X-ray source
($L_{2-10\,\rm keV} > 10^{41}\rm erg~s^{-1}$).

We find a photospheric temperature for this source of $\tch\sim
110\rm~K$ and an outer core temperature of $T_{c}=14\rm~K$ for a density
profile of $\krho=2$ at a confidence level of 43 \%.  
If we allow the density profile to vary, we find
that $\krho$ is well constrained, with values outside
$1.5 \la \krho \la 2.1$ 
being statistically ruled out.
We infer a luminosity of $7\times 10^{11}L_{\odot}$, a size $R_c\simeq 5$~kpc, and a
characteristic radius $\rch\simeq 34 \rm ~pc$.  As always, our estimate of the
total size includes any extended, optically thin emission.
Soifer et al. (2000) found that at mid-infrared ($7.9-24.5 ~\micron$)
wavelengths, the diameter of this source is less than
$0.25''~(205\rm~pc)$.  
This is in good agreement with our estimate of the characteristic emission radius
at 24 \micron, $r_m(24\;\mu{\rm m})\simeq 220~\rm pc$.
This is larger than the characteristic radius (34 pc) since the opacity at 24 \micron\ is significantly
larger than it is at the
characteristic frequency, which corresponds to $\lambda\sim 100~\micron$).
From our inferred SED parameters, we find that this
source has $\Sigma\delta \simeq 0.07 \rm g~cm^{-2}$ and $L/M\delta
\simeq 26\;L_\odot/M_\odot$.
In contrast to Arp 220, UGC 5101 has $F_{100\mu\rm m} \ga F_{60\mu\rm m}$.
We infer that it has a lower value of $\rct$ ($\sim 160$ vs. 370) and a higher  
mean surface density ($\Sigma \sim 0.07~\rm g\;cm^{-2}$ vs. 0.025~g~cm$^{-2}$).

IRAS 08572+3915 differs from the previous two sources in having
$F_{100 \mu\rm m} < F_{60 \mu\rm m}$.
This source appears to contain an AGN
based on an analysis of PAH features (Imanishi \& Dudley 2000) and on
the compactness seen from imaging (Surace et al 1998, Soifer et al
2000).   It is the warmest ($\tch \sim
170 \rm K$) source in our sample, with $L/M\delta
=128~L_{\odot}/M_{\odot}$,
and the lowest value of the surface
density, $\Sigma\delta\sim 0.01 \rm ~g~cm^{-2}$.  As previously, we have
solved for these parameters simultaneously, along with the density
variation, which yields a best-fit density profile of $\krho=2$ at a
confidence level of 83 \%.  Density profiles greater than 2.1 and less
than 1.7 are not favored statistically.  

Density profiles of $\krho=2$ are supported by the resolved observations
of DS98 for other sources in our sample as well.  
Furthermore, our mass estimates based on dust emission are in 
good agreement with those of DS98, which are based on CO emission.
As noted earlier, 
our estimate of the mass for Arp 220 is in excellent agreement with
that measured by DS98.
Resolved observations of Mrk 231 yield a gas mass very similar to ours; they estimate a gas
mass interior to 1.7 kpc of $4 \times 10^{9} M_{\odot}$, which is within
a factor of 1.5 of our estimate for $r < 1.7 \;\rm kpc$.  DS98 cite
a mass for IRAS 10565+2448 of $4\times 10^{9}~M_{\odot}$ for $r<1.6~\rm kpc$ 
which is within a factor of 1.5 of our value for the mass enclosed interior
to that radius. 
On average, our estimates of the mass (based on $\delta=1$) are about 1.3 times 
those of DS98, which is excellent agreement in view of the uncertainties in
each method and of possible variations in the value of $\delta$ from galaxy to galaxy.

\section{Inferred Parameters for Sub-millimeter Galaxies}

We present here our SED fits and inferred parameters for several SMGs.
We focus our discussion of the SEDs of SMGs on the observations of
 Kovacs et al. (2006), which make use of SHARC-2 $350~\micron$ measurements; these are
currently the most direct probe of the rest-frame FIR for $z\sim 2$ systems.
The SEDs of these sources are shown in Figures 8-12 and the source
and SED parameters are reported in Table 2.  For sources for which there
are no data shortwards of the peak, we set bounds on $\rct$ guided by
our experience with local ULIRGs, i.e., specifically we set the minimum 
and maximum values to be 40 and 600, respectively, which encompasses the range 
of values of $\rct$ that we found for local ULIRGs.  
We take $\krho=2$ for all of these sources,
as we found that $\krho=2$ provided the best fit to the data of local ULIRGs
and is independently supported by
CO observations of DS98.

Recall that
the break frequency, $\nu_{\rm break}$, is the frequency at which the emission changes in
character from being dominated by the cool material on the outside to
warmer material that is deeper inside the envelope. For the
ULIRGs listed in Table 1, the typical break frequency is $\nu_{\rm break}\simeq 1 \times 10^{12} ~\rm Hz$.
However, note that moving a local ULIRG to
higher redshift moves the break frequency to lower frequencies by a factor
of $(1+z)$---which means that even mm wavelengths can be in the
intermediate frequency region for high-redshift systems such as SMGs.  
Normalizing to a typical value of $\rct\sim 100$, we find from
equation (\ref{eq:nubreak1}) that
\beq 
\nu_{\rm break}=6.35\times 10^9\left(\frac{100}{\rct}\right)^{0.4}\frac{\tch}{1+z} \; ~~
\mbox{Hz},
\label{eq:nubreak2} 
\eeq 
for a typical value of the density profile, $k_{\rho}=2$.
For frequencies less than this break frequency, the slope of the spectrum
transitions to the low frequency (Rayleigh-Jeans) regime and one may
use equation (\ref{eq:mass_inf}) to solve for the mass given the
millimeter flux.

For the source SMMJ163658.19+410523.8 from Kovacs et al. (2006)
(abbreviated as SMMJ1636581 in Table 2), which is at $z=2.454$, we estimate
a luminosity in the range of $6.8 -8.2 \times 10^{12} L_{\odot}$, which is 
close to the estimate of $8.5 \times 10^{12} L_{\odot}$ cited by Kovacs et al. (2006), which they obtained by fitting a modified greybody to the observed
SED.    
The range of values for which 
the chi-squared per degree of freedom changes by less than unity for $\rct$ and $\tch$ are $40-150$ and $103-115~\rm K$, respectively.  The large uncertainties in the determination
of the SED quantities and the corresponding source parameters, particularly the surface density 
are due to 
this source having a rest-frame SED that peaks at a wavelength shorter than 
the observed $350~\micron$ band.  $\it{Herschel}$ PACS observations 
(the $170~\micron$ band) are needed to firmly constrain the SED and source 
parameters of this galaxy.

We find that the source SMMJ13650.43+405737.5,
which is at $z=2.378$
has a luminosity of $5\times 10^{12} L_{\odot}$,
also quite close to the value cited by Kovacs et al. (2006).  This source peaks at a longer
wavelength, and SHARC-2 observations are sufficient to more firmly constrain the SED and
source parameters.  Finally, the source SMMJ105238, which is the highest redshift SMG 
among these sources ($z_{\rm spec}=3.036$),
has a rest-frame SED that peaks at the shortest wavelength, and therefore has the highest $\tch$ of any source
in this sample.  Our inferred size, 13.4 kpc,
is also larger than what we found for the 
other sources.  

Due to the small number of rest-frame FIR observations
of $z \sim 2$ SMGs, our sample of galaxies studied here is necessarily limited.  
Nonetheless, there are several trends that do stand out.
The SMGs we have studied have higher luminosities than local ULIRGs
by a factor of 5.5 on average, i.e., the luminosities
are all in excess of $4 \times 10^{12} L_{\odot}$.
The gas masses are also higher than local ULIRGs by a factor of $\sim 8$,  
but the 
geometric mean luminosity-to-mass ratio of these galaxies
is $\simeq 40$,
 which is within a factor 1.4 of that of local ULIRGs (see below); 
the scatter in $L/M$ about this mean is less than a factor 2.

\section{FIR Photometric Redshifts}

Chapman et al. (2003, 2005) have been very successful at obtaining spectroscopic 
redshifts of SMGs from optical measurements, guided by radio or optical
associations.   This approach has yielded about a hundred accurate spectroscopic
redshifts.  However, spectroscopic redshifts
may not be available for the large samples of SMGs that are expected to be
observed by upcoming FIR instruments like $\it{Herschel}$.  
We present here a means of inferring the redshifts of 
dusty galaxies from FIR photometric data.  We present the derivation
of this method both using our formalism and using very simple relations that
are independent of the details of our formalism.

The parameters we infer, $L/M$ and $\Sigma$, depend on redshift
through the dependence of frequency on redshift.  We can express the
redshift dependence of these parameters in a very simple manner.  
Since the luminosity of a dust envelope satisfies $L \propto R^{2} T^{4}$
and the inferred mass is $M \propto \Sigma R^{2}$,
we have: 
\beq
\frac{L}{M} \propto \frac{T^{4}}{\Sigma}  \; .
\eeq
The redshift dependence of the surface density, $\Sigma$, follows
from noting that the optical depth at the observed frequency
must match that at the emitted frequency, since $\tau_\nu$ determines
the shape of the SED, which is invariant: $\kappa(\nu_\obs)\Sigma_\obs
=\kappa(\nu_\rest)\Sigma_\rest$, so that $\Sigma_\obs/\Sigma_\rest=(\nu_\rest/\nu_\obs)^\beta=
(1+z)^\beta$.
Since $T_{\rm obs}=T_{\rm rest}/(1+z)$, it follows that
\beq
\left(\frac{L}{M\delta}\right)_{\rm obs} = \left(\frac{L}{M\delta}\right)_{\rm rest} (1+z)^{-(4+\beta)}\; ,
\label{eq:LM_z_easy}
\eeq
where we have included the dust-to-gas parameter $\delta$ to emphasize that
we are actually determining the dust mass.

Alternatively, we can derive the redshift dependence of $L/M$ within the context of our formalism.
Using $\nu_{\rm ch,\, obs}=\nu_{\rm ch,\,rest}/(1+z)$ in equations (\ref{eq:LM}) and
(\ref{eq:Sigma}), and noting that $\rct$ is independent of redshift
since it is a dimensionless ratio of two lengths, it is
straightforward to show that: 
\beq \Sigma_{\rm
obs}=\Sigma_\rest(1+z)^{\beta},~~~~~~\left(\frac{L}{M\delta}\right)_{\rm
obs}=\left(\frac{L}{M\delta}\right)_\rest(1+z)^{-(4+\beta)}\; .
\label{eq:sigma_LM_z} 
\eeq 
Figure 6 shows the change in these
parameters, when a local ULIRG $(z\ll 1)$, with inferred parameters
similar to Arp 220, is moved to $z=1$.  Note that the line along which the
source is moving as a function of redshift is a line of constant
$\rct$.  The parameters reported in Table 1 (and Table 2 for SMGs) are the intrinsic
parameters; observed parameters can be obtained by using the relations
in equation (\ref{eq:sigma_LM_z}) along with the redshifts cited for
the sources.

We can now use
equation (\ref{eq:sigma_LM_z}) or (\ref{eq:LM_z_easy}) to infer the redshift of a dusty
galaxy from its observed value of $L/M\delta$,
which we have shown can be derived from the FIR SED analytically: 
\beq 1+z_{\rm inf}=\left[\frac{\langle L/M\delta\rangle}{\left(L/M\delta\right)_{\rm obs}}\right]^{1/6}\; ,
\label{eq:zdet}
\eeq 
where $\langle L/M\delta\rangle$ is the typical value; since $L/M\delta$ can range over
an order of magnitude, we use the geometric mean. 
Note that $L$ is the FIR luminosity, which is determined by our analytic fit to data
at rest wavelengths $\ga 60$~\micron.
Here we have taken advantage
of the result that $\beta\simeq 2$.
For $L/M\delta$ within a factor 3 of the mean, the uncertainty in $1+z$ is only $3^{1/6}=1.20$.

The geometric mean of the intrinsic $L/M\delta$ values of the ULIRGs in Table 1 is
$\langle L/M\delta\rangle = 60
 ~L_{\odot}/M_{\odot}$;
 all the ULIRGs in our sample have $L/M\delta$ within a factor 3 of this. The inferred
 redshift for a ULIRG at high redshift is then
 \beq 
 1+z_{\rm inf}=\left[\frac{60}{\left(L/M\delta\right)_{\rm obs}}\right]^{1/6} ~~~~~~
\mbox{(ULIRG normalization)}
\label{eq:zdetu}
\eeq 
As we show below, the typical uncertainty for deriving redshifts for SMGs using
this normalization  is  $\sim 10\%$. 

To demonstrate the applicability of our redshift inference method, we first 
infer the redshift for a test case where we know the redshift exactly.  We place 
the observed SED of Arp 220 at a range of redshifts, from $z=0.018-10$ 
(Fig  7a), 
and use our method to infer the redshift.
To test the sensitivity of our method to redshift, we first replace
$\langle L/M\delta\rangle$ with the value for Arp 220 
and infer $(1+z_{\rm inf})/(1+z)$,
where $z_{\rm inf}$ refers to our inferred redshift and $z$ is the actual redshift (Fig 7b). 
We find that $0.95<(1+z_{\rm inf})/(1+z)<1$,
which demonstrates that in principle our method should work out to redshifts $z\sim 10$.
Next, we use our method with the mean value of $L/M\delta$ (eq. \ref{eq:zdetu}); since
Arp 220 has a relatively low light-to-mass ratio ($L/M\delta\simeq 25$),
the inferred redshift is somewhat higher than the actual one, $1.09<(1+z_{\rm inf})/(1+z)<1.15$.

We now infer the redshifts for a sample of SMGs at $z \sim 2-3$ studied recently by 
Kovacs et al. (2006) using SHARC-2 $350~\micron$ measurements (which
probe close to the rest-frame FIR for $z\sim 2$ galaxies), and
SCUBA and MAMBO measurements.  
We  illustrate our method with
two different normalizations. Prior to the determination of spectroscopic
redshifts for the SMGs, they would most likely have been compared with
local ULIRGs; their FIR photometric redshifts could have then been
determined from equation (\ref{eq:zdetu}). Now that redshifts have been measured
for some of the SMGs, it is also possible to use a normalization appropriate for them:
\beq 
 1+z_{\rm inf}=\left[\frac{40}{\left(L/M\delta\right)_{\rm obs}}\right]^{1/6} ~~~~~~
\mbox{(SMG normalization)}\; .
\label{eq:zdets}
\eeq 
Note that once a larger sample of SMGs with measured SEDs and
spectroscopic redshifts becomes available,
it will be possible to improve the accuracy of  the average $L/M\delta$ for SMGs in this equation.
 We compare our estimated redshifts using both normalizations
 with the measured spectroscopic
redshifts in Table 3.  We define the accuracy of 
the redshift determination as
\beq
A\equiv \frac{\mbox{Max}(1+z,1+z_{\rm inf})}{\mbox{Min}(1+z,1+z_{\rm inf})}\; .
\eeq
The average accuracy of the redshifts using the ULIRG normalization
for the sources in Table 3 is $\langle A\rangle=1.11$. 
Using the SMG normalization, the average accuracy improves to 1.05.

An alternate method of inferring redshifts utilizing FIR photometry is
to infer the redshift from the longwards shift in the observed-frame peak of the SED
(e.g., Chapman et al. 2005).  
There are two problems with this: First, it is difficult to determine the peak frequency accurately
from poorly sampled, noisy data. Second,
the intrinsic range of values of $\tch$ 
is larger than the range of values of $(L/M\delta)^{1/6}$. For the sample of ULIRGs in Table 1, 
$\tch$ varies by a factor 1.3 from the average value of $132$, whereas
$(L/M\delta)^{1/6}$ varies only by a factor 1.20. For the sample of SMGs in Tables 2 and
3, $\tch$ varies by a factor 1.22 from the mean, whereas the inferred redshifts
from our method are accurate to within a factor 1.15 and 1.08 (at worst; the typical
accuracies are 1.1 and 1.05 respectively) using the
ULIRG normalization and the SMG normalization, respectively. Thus, in each case
the $L/M\delta$ method gives a value of $1+z$ that is more accurate by a factor of about
1.5.

We note that an important caveat to applying our redshift
estimation method is that normal galaxies at low redshifts that have
lower star formation rates than ULIRGs and have intrinsically
lower $L/M\delta$ values can be misidentified as high redshift ULIRGs, if
only FIR data are available.  This problem can be remedied through 
the use of multi-wavelength photometry.  Dusty ULIRGs radiate most of their energy in the infrared,
so intrinsically low $L/M\delta$ galaxies can be identified
by comparing the near-IR (or UV/optical) photometry with the 
FIR, so as to select only those where the latter dominates. 

The Infrared Array Camera (IRAC) on board the $\it{Spitzer~Space~Telescope}$ has 
been instrumental in obtaining photometric
redshifts (Brodwin et al. 2006) of high redshift galaxies. This method primarily utilizes the approximate
constancy of the rest-frame stellar peak at $\lambda=1.6~\micron$ (Simpson 
\& Eisenhard 1999; Sawicki 2002).  Although near-IR photometric redshift
codes are highly successful and sophisticated, utilizing template fitting
Monte Carlo 
algorithms or neural networks, a key component of these
various algorithms is to sample the $1.6~\micron$ stellar peak, which 
cannot be accomplished for $z \ga 4$ by IRAC.  Moreover, these codes
generally have difficulty in obtaining robust photometric redshifts
for power-law AGN-type near-IR SEDs, where the stellar bump is 
weak or
absent.
As such, many photometric redshift codes calibrate their methods on
SED samples that exclude AGN-type SEDs.
By contrast, several of the sources in our local ULIRG sample are optically 
classified quasars, such as Mrk 1014 and Mrk 231, while others such as IRAS 08572+3915
are inferred to have energetically active AGN on the basis of hard X-ray measurements.
Our FIR photometric redshift method may be useful as a complementary
technique for sources with AGN.
These sources do have somewhat larger values of $L/M\delta$ (by about a factor of 2 relative
to the geometric mean of $L/M\delta$).  It is possible that when larger sets of
FIR observations of high redshift ULIRGs become available with $\it{Herschel}$ and
SCUBA-2 observations, one may be able to robustly identify two 
subclasses of sources, corresponding to galaxies with energetically active AGN ($\sim 2$ times higher $L/M\delta$)
and those without energetically active AGN.  Inferring photometric redshifts within these 
subclasses with the appropriate $L/M\delta$ may further improve the accuracy.

\section{Conclusions}

We have applied the shell methodology for radiative transfer developed
in CM05 to a
range of extragalactic sources, from local ULIRGs to high redshift
SMGs.  The main results are:

1. Using the general expressions for the SEDs of dusty sources given in CM05,
we have shown how to derive the light-to-mass ratio,
$L/M\delta$, and the mean surface density, $\Sigma\delta$,
of dusty galaxies at cosmological distances.  Here $M$ and $\Sigma$ refer to the gas mass,
and $L$ is the FIR luminosity as determined from observations at rest wavelengths $\ga 60$~\micron.
The effective dust-to-gas ratio, $\delta$, is the ratio of the actual FIR opacity
to the one we have adopted, which is twice the Weingartner \& Draine (2001) opacity;
the factor 2 allows for ice mantles.
Approximate expressions are given for the case in which the radius of the dusty envelope,
$R_c$, is much larger than the characteristic photospheric radius, $\rch$
(i.e., for $\rct\equiv R_c/\rch\gg 1$). 

2.  The long-wavelength slope of ULIRGs can be fit with
a standard dust opacity curve ($\kappa_\nu\propto \nu^2$ in
the FIR) when a self-consistent radiative
transfer solution is employed.  This confirms the conclusion of Dunne \& Eales (2001),
who used two-temperature fits to the SED.
Comparison with the three 
sources in DS98 for which there are resolved measurements for the galaxies
in our sample (Arp 220, Mrk 231 and IRAS 10565+2448) shows that on average our
mass estimates are about 1.3 times greater than theirs, which is excellent agreement
in view of the approximations in each method and the possible variations in
the value of $\delta$ from galaxy to galaxy.

3. From an analysis 
of the FIR data of 10 local
ULIRGs, we find that the mm to FIR SED can be well-described
by our simple construct of a spherically symmetric dust envelope
surrounding a central source of luminosity.  We find that a 
density profile $\rho \propto r^{-2}$ provides the best to the FIR data and is
consistent with CO masses.  We report our findings for the luminosities, masses, and
sizes of these 10 ULIRGs.

4.  We find that local ULIRGs and high redshift SMGs ($z \sim 2$) have
similar $L/M\delta$ ratios, with the SMGs in our sample having values
$\sim 1.45$ times smaller. The SMGs in our sample have luminosities about 5.5 times larger,
and masses about 8 times larger, than the ULIRGs in our sample.

5.  We have developed a method of inferring FIR photometric redshifts.
The accuracy of the method depends on whether the galaxy has
a light-to-mass ratio comparable to the template, but since $1+z$
scales as only the 1/6 power of $L/M\delta$, significant variations are allowed.
Using Arp 220 as an example, we showed that our method should work for
redshifts $z\la 10$.
We tested our method on a sample of five SMGs for which good photometric
data are available. Under the assumption that the SMGs had the same
light-to-mass ratio as a sample of local ULIRGs, we were able to infer values of
$1+z$ for the SMGs with an average accuracy of 10\% and a worst accuracy
of 15\%. If we used the average light-to-mass ratio for the SMGs, the
average accuracy improved to about 5\%. As the samples of high-redshift galaxies
grow, our knowledge of both the typical light-to-mass ratio and the accuracy of
FIR photometric redshifts will improve.
Our method should be applied to
galaxies that radiate most of their energy in the FIR, such
as ULIRGs and SMGs.  Whether this condition is met can be determined by 
confirming that
the near-IR (or optical/UV) luminosity is significantly less than
that emerging in the FIR.  
Our method works even if AGN provide a significant fraction of
the luminosity, although our sample is not large enough
to determine how large the AGN fraction can become before
our method breaks down.
As our method is analytic, it can be 
employed to quickly obtain photometric
redshifts of large samples of SMGs, as are expected to be detected in
the FIR by the upcoming $\it{Herschel}$ mission.  This
FIR photometric redshift method provides a complementary
means of inferring the redshift when near-IR methods are
not available or are not viable.

6. We discuss how our approximation for the radiative transfer
compares with the standard single-temperature SED in Appendix B.
The accuracy of the single-temperature blackbody approximation degrades
for extended envelopes, $\rct \ga 100$, but
the approximation is typically accurate to within a factor 2 for compact envelopes, for
which the source function does not probe a large range of
temperatures.  The effective dust temperature in the single-temperature approximation is close
to the outer core temperature. 

\bigskip
\bigskip
\acknowledgments

We thank Henrik Beuther, Jack Welch, Carl Heiles, Phil Myers, 
Jonathan Tan, Susana Lizano, Dave Sanders, Yeshe Fenner, and Ski
Antonucci for helpful discussions.  We especially thank Erik
Rosolowsky for many helpful and informative discussions
on the inference of gas masses and CO observations.  
The research of CFM is supported
in part by the NSF through grants 
AST06-06831, and 
PHY05-51164.
The research of SC is supported by a
National Science Foundation postdoctoral fellowship.

\appendix
\section{Analytic SEDs}

Here we summarize the analytic form for the mm to far-IR SED derived in CM05:
\begin{equation} L_{\nu}=16\pi^2(k_{\rho}-1)
\rch^{2}\left(\frac{2h\nu_{ch}^{3}}{c^{2}}\right)
\tilde{\kappa}_{\nu}\tilde{\nu}^{3}\tilde{r}_{m}^{2-k_{\rho}}
\Delta\tilde{r}_{m}\exp\left[-\frac{h\nu}{kT(\tilde{r}_{m})}-
\tau_\nu(\tilde{r}_{m})\right]\; .
\label{eq:shell}
\end{equation} The optical depth, $\tau_{\nu}$ is given by: \beq
\tau_{\nu}=\tilde{\kappa}_{\nu}\left(\tilde{r}^{-k_{\rho}+1}-
\rct^{-k_{\rho}+1}\right) \; .  \eeq The temperature profile,
$T(\tilde{r}_{m})$, is given by $T=\tch \tilde{r}^{-k_{T}}$,
with $k_{T}$ and $\tilde{r}_{m}$ specified below.

The characteristic normalized emission radius, $\tilde{r}_{m}(\nu)$, is the
location in the shell where most of the flux in a given frequency-band
originates from, (the ``$m$'' is for maximum), and is given by
\beq
\tilde{r}_{m}={\rm Min}\left (\tilde{r}_{\rm m,low-int}+\tilde{r}_{\rm
m,high},\rct\right)\; ,
\label{eq:rmp} 
\eeq 
where the total $\tilde r_{m}$ is the sum of the high
frequency $\tilde r_{m}$ and the combined low-intermediate frequency $\tilde r_{m}$,
\beq 
\tilde{r}_{\rm m,high}=\left[\frac{\tilde\kappa_\nu
(k_\rho-1)}{\tilde\nu k_T}\right]^{1/(k_T+k_\rho-1)}\; , \eeq \beq
\tilde{r}_{\rm m,low-int}=\frac{\rct
C^{1/\kt}}{\rct\tilde{\nu}^{1/k_{T}}+C^{1/\kt}}\; .  
\eeq 
The parameter $C$ is the ratio of the typical value
$h\nu$ to $kT$  in the intermediate frequency regime, and is given by
 \beq
C=0.3+1.5k_{\rho}-0.78k_{\rho}^{2} \; .
\eeq 
The power law for the temperature profile is approximately
\beq
k_{T}=\frac{0.48k_{\rho}^{0.05}}{\rct^{0.02k_{\rho}^{1.09}}}
+\frac{0.1k_{\rho}^{5.5}}{\rct^{0.7k_{\rho}^{1.9}}}\; .  
\eeq
The preceding two 
relations hold for $1\la \krho\la 2$ 
and $\rct\ga 2$
to within $\sim$ 10 \% accuracy.

The shell thickness, $\Delta\tilde{r}_{m}$ is given by: 
\beqa 
\lefteqn{
\Delta\tilde{r}_{m}=\left[\frac{\Gamma\zeta \exp\left(\nut
\rt_{m,\,\rm low-int}^\kt\right) }
{(3-k_{\rho}-k_{T})\tilde{\nu}\Gamma\zeta\rct^{-(3-k_{\rho}-k_{T})}+k_{T}
\tilde{\nu}^{(3-k_{\rho})/k_{T}}}\right]\frac{1} {\tilde{r}_{m,\,\rm
low-int} ^{2-k_\rho}}}\nonumber \hspace{3cm}\\
&&+\frac{(2\pi/h_{m}'')^{1/2}}{1+2\tilde{\kappa}_{\nu}
\left(\frac{k_{\rho}-1}{k_{\rho}+1}\right)(\tilde{r}_{m,\,\rm high}
^{1-k_{\rho}}- \tilde{r}_{m,\,\rm high}^{2}\rct^{-k_{\rho}-1})}\; ,
\eeqa 
where the argument $(3-k_{\rho})/k_{T}$ for the Gamma and Zeta
functions has been suppressed for clarity and where
\beq
h_{m}''=\tilde{\nu}\kt(\kt-1)\rt_{\rm m,high}^{\kt-2}+\krho(\krho-1)\tilde{\kappa_{\nu}}\rt_{\rm m,high}^{-\krho-1}  \; .
\eeq

\section{ Inference of Masses from the Low-frequency
Dust Continuum}

It is possible to infer the mass of gas in a source of known redshift
directly from the observed flux
and quantities that describe the SED.
At low frequencies ($\nu<\nu_{\rm break}$), the source is both optically
thin and the temperature is high enough that the emitted radiation is
in the Rayleigh-Jeans regime. As a result,
the spectral luminosity is
\beq
L_\nu=\int_{R_d}^{R_c}(4\pi \rho\kappa_\nu B_\nu)4\pi r^2 dr\; ,
\eeq
where the Planck function $B_\nu=2kT/\lambda^2$ in the Rayleigh-Jeans regime.
Dust sublimates inside the dust destruction radius $R_d$, and we assume that
this is negligible compared to the size of the source, $R_c$.
For power-law density and temperature profiles, $\rho=\rho_{\rm ch}\rt^{-\krho}$
and $T=\tch \rt^{-k_T}$, we then find
\beq
L_\nu=4\pi \rho_{\rm ch} \kappa_\nu\left(\frac{2 k\tch \nu^2}{c^2}\right)4\pi\rch^3
\left(\frac{\rct^{3-\krho-k_T}}{3-\krho-k_T}\right)\; .
\eeq
This result also follows directly from equations (4), (12) and (16) in CM05.
The mass of gas producing the emission is
\beq
M=\int_0^{R_c} 4\pi r^2\rho_\ch \rt^{-\krho} dr=
4\pi \rho_{\rm ch}\rch^{3}\left(\frac{\rct^{3-\krho}}{3-\krho}\right) \; .
\eeq
Noting that the temperature at the outer edge is
$T_c=\tch\rct^{-k_T}$,
we can relate the mass to the luminosity,
\beq
M=\left(1-\frac{k_{T}}{3-k_{\rho}}\right)
\left(\frac{c^{2}}{8\pi k T_c}\right)
\frac{L_{\nu}}{\nu^2\kappa_{\nu}}\; .
\label{eq:mass_inf_L}
\eeq
To this point, all frequencies are measured in the rest frame of the source.
Converting to observed frequencies and using equation
(\ref{eq:rest}), we can relate the mass to the observed flux,
\beq
M=\left(1-\frac{k_{T}}{3-k_{\rho}}\right)
\left(\frac{c^2}{2 kT_{c,\, \rest}}\right)
\frac{D_L^2F_{\nu\,\obss}}{(1+z)^3\nu_\obss^2\kappa[(1+z)\nu_\obss]}\; ,
~~~~~~~([\nu_\obss(1+z)<\nu_{\rm break}],
\label{eq:mass_inf}
\eeq
where $\nu_{\rm break}$ is given by equation (\ref{eq:nubreak1})
or, for the case $\krho=2$, by equation (\ref{eq:nubreak2}). 
Note that this expression for the mass depends on only
one parameter from the SED, the temperature at
the edge of the core, $T_{c,\,\rest}=T_{\rm ch,\,\rest}\rct^{-k_{T}}$.

One of the standard approximations used in the literature to infer
source parameters from SEDs is Hildebrand's (1983) approximation,
where one assumes an isothermal distribution of dust.  To illustrate
the difference between our solution,
which self-consistently takes into account the temperature 
variation in the envelope,
and the standard Hildebrand
solution, we over-plot our solution with DUSTY's numerical results for
a large $\rct$ case, $\rct\sim 300$, and a small $\rct$ case,
$\rct\sim 10$, in Figures 13 and 14.  As is clear, the accuracy of such
single-temperature fits, relative to the numerical solution, degrades
for large $\rct$, but is nearly as accurate as our
solution for low $\rct$.  Similarly, the inference of source
parameters using Hildebrand's prescription also degrades at large
$\rct$ to a factor of $\sim 2$, while our solution is accurate (for
this large $\rct$ example) to within 10 \%.

If we wanted to characterize the envelope as having a single
temperature, we can see from equation (\ref{eq:mass_inf_L}) that that
equivalent temperature is
\beq 
T_{\rm dust}=\left(1-\frac{k_{T}}{3-k_{\rho}}\right)^{-1}T_{c} \; , 
\eeq
where for the remainder of this discussion we shall assume $z=0$ and
where $T_{\rm dust}$ is now the $\it{single}$ temperature
characterizing the entire envelope, and not the photospheric
temperature, which has a particular meaning (see CM05).  However, this
is not the temperature that one uses when one uses the Hildebrand
approximation (or single-temperature blackbody approximation).  
Instead, one uses the frequency at which the spectrum peaks, which
for a modified blackbody is
$h\nu_{\rm peak}=(3+\beta_{\rm iso})kT$, 
where $\beta_{\rm iso}$ is the value of the opacity index used in the
isothermal fit.
For an isothermal dust distribution ($k_T=0$ and $T=T_{\rm dust}$),
we can rewrite equation (\ref{eq:mass_inf_L}) in terms of the peak frequency as 
\beq 
M_{\rm iso}=
\left[\frac{(3+\beta_{\rm iso})c^2}{8\pi h\nu_{\rm peak}}\right]
\frac{L_{\nu}}{\nu^2\kappa_\nu}\; .
\eeq 
The value of $\beta_{\rm iso}$ used in single temperature
solutions is often $\sim 1$ (Yun \& Carilli 2002, Dunne \& Eales
2000).

To understand why the
Hildebrand approximation deviates from our solution (and the numerical
solution) at large $\rct$,
we express the ratio of our mass estimate
to the isothermal estimate as
\beq
\frac{M}{M_{\rm iso}}=
\left(1-\frac{k_{T}}{3-k_{\rho}}\right)\frac{\tilde\nu_{\rm peak}\rct^{k_{T}}}
{\left(3+\beta_{\rm iso}\right)} \; ,
\label{eq:mass_cond} 
\eeq 
where we used the relation $h\nu_{\rm peak}/kT_c=h\nu_{\rm peak}\rct^{k_T}/k\tch
=\tilde\nu_{\rm peak}\rct^{k_T}$.
In general, the peak frequency is a function of $\rct$, as we have
discussed in CM05; for $\krho=2$ and large $\rct$, it approaches
1.64 (eq. \ref{eq:nupk3}). For large $\rct$, we have $k_T\simeq 0.4$,
which for $\beta_{\rm iso}=1$
implies that $M/M_{\rm iso}\simeq (\rct/30)^{0.4}$.
Hence, the isothermal approximation becomes increasingly inaccurate
as $\rct$ increases. This is to be expected, since emission arises from
a wide range of radii when the envelope is very distended, with $R_c \gg \rch$.

\clearpage

\begin{figure}[!ht] \begin{center}
\centerline{\psfig{file=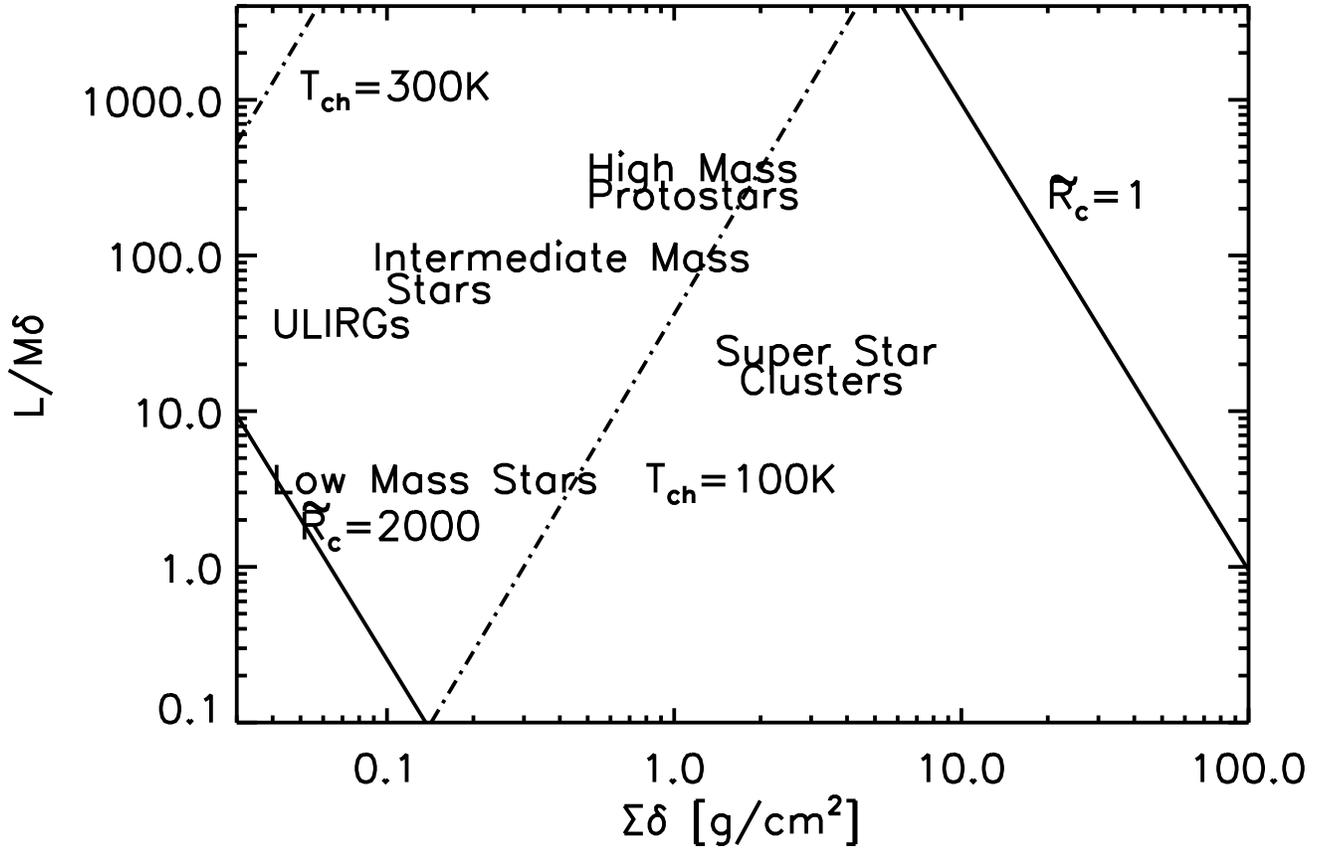}}
\end{center}
\caption{$L/M\delta$ vs $\Sigma\delta$ plot for density profile,
$k_{\rho}=3/2$ and dust model, Weingartner \& Draine (2001) coated
with ice mantles}
\end{figure}

\begin{figure}[!h] \begin{center}
\centerline{\psfig{file=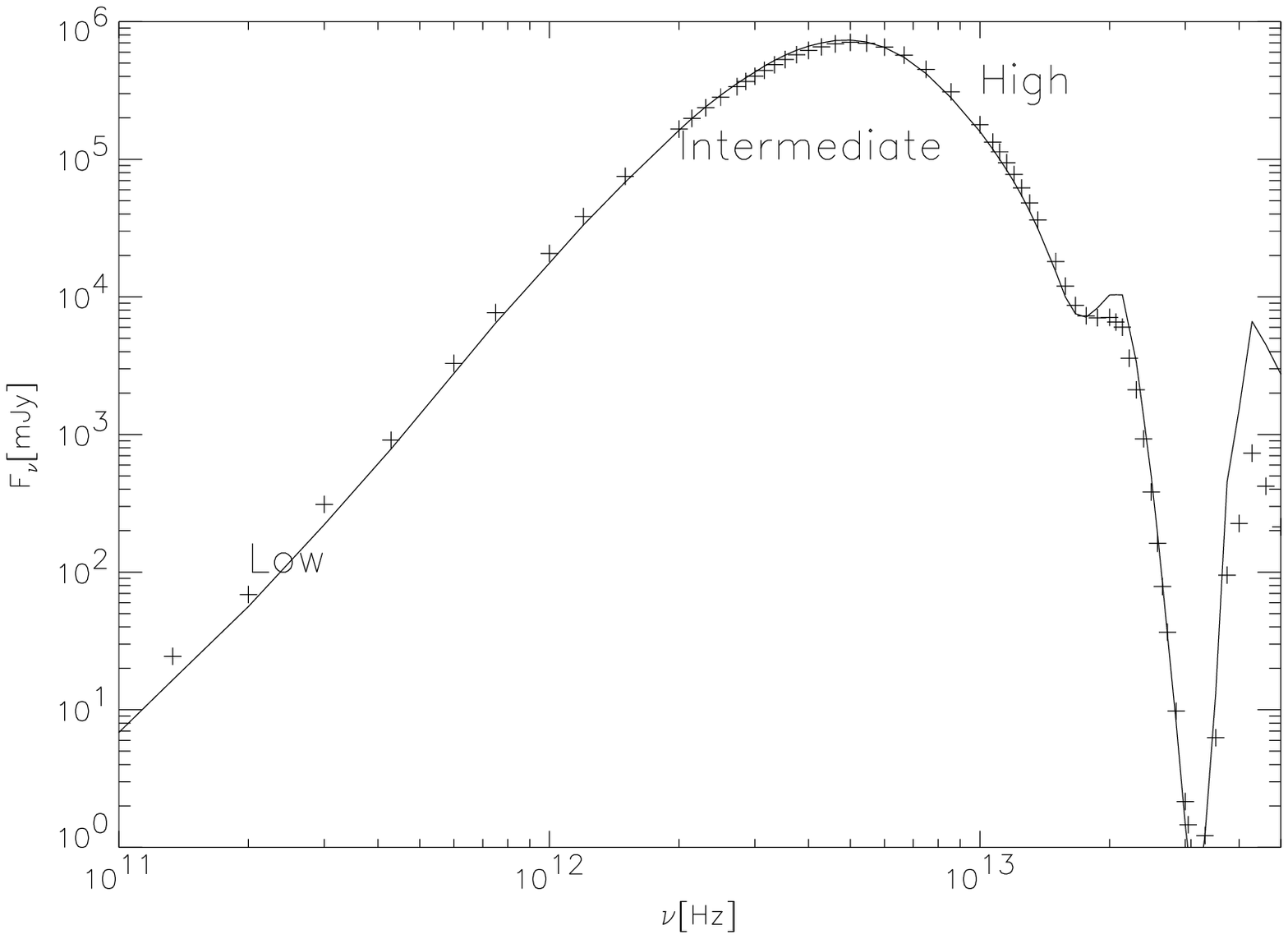,height=2.8in,width=2.8in}
\psfig{file=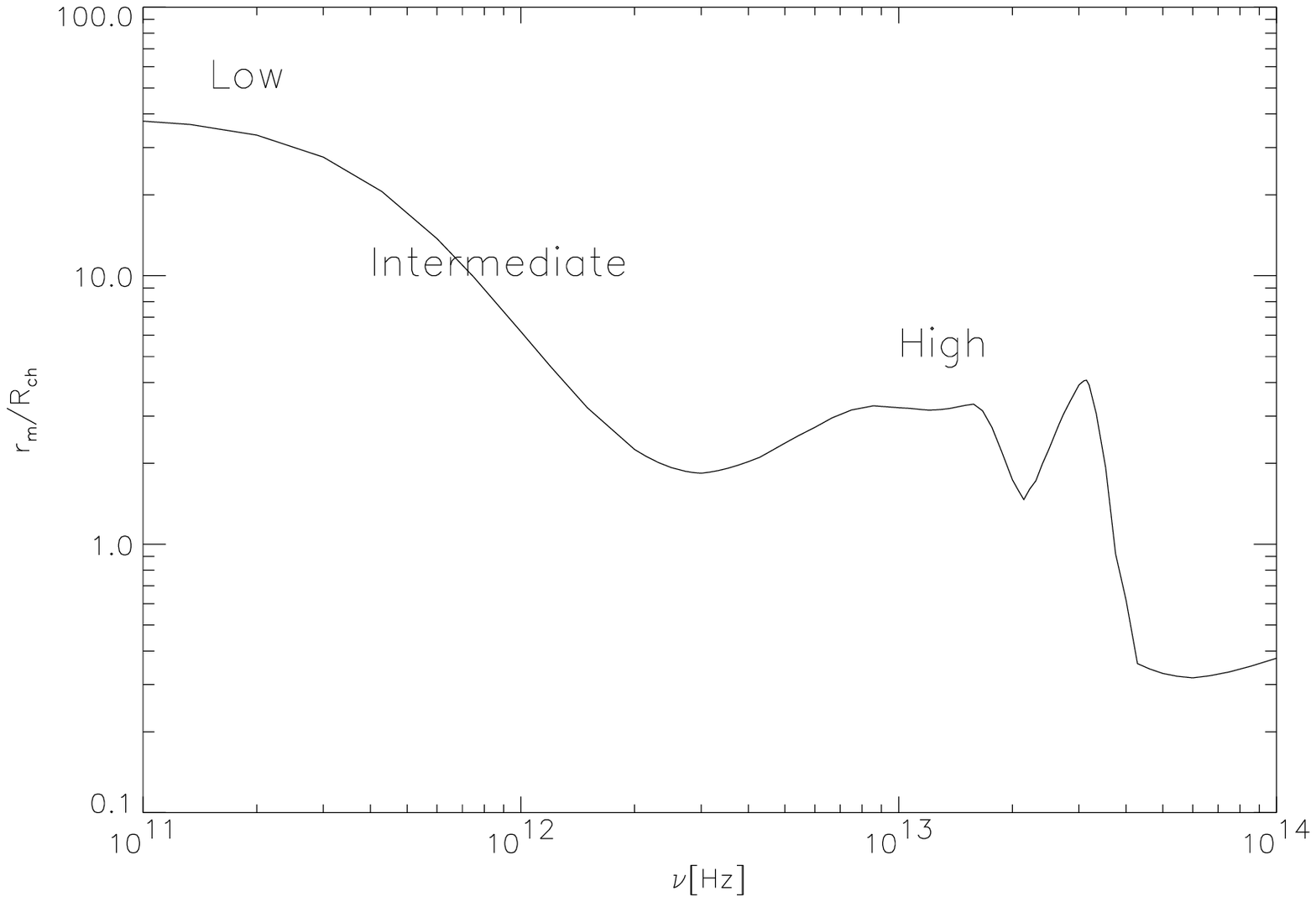,height=2.8in,width=2.8in}
{\psfig{file=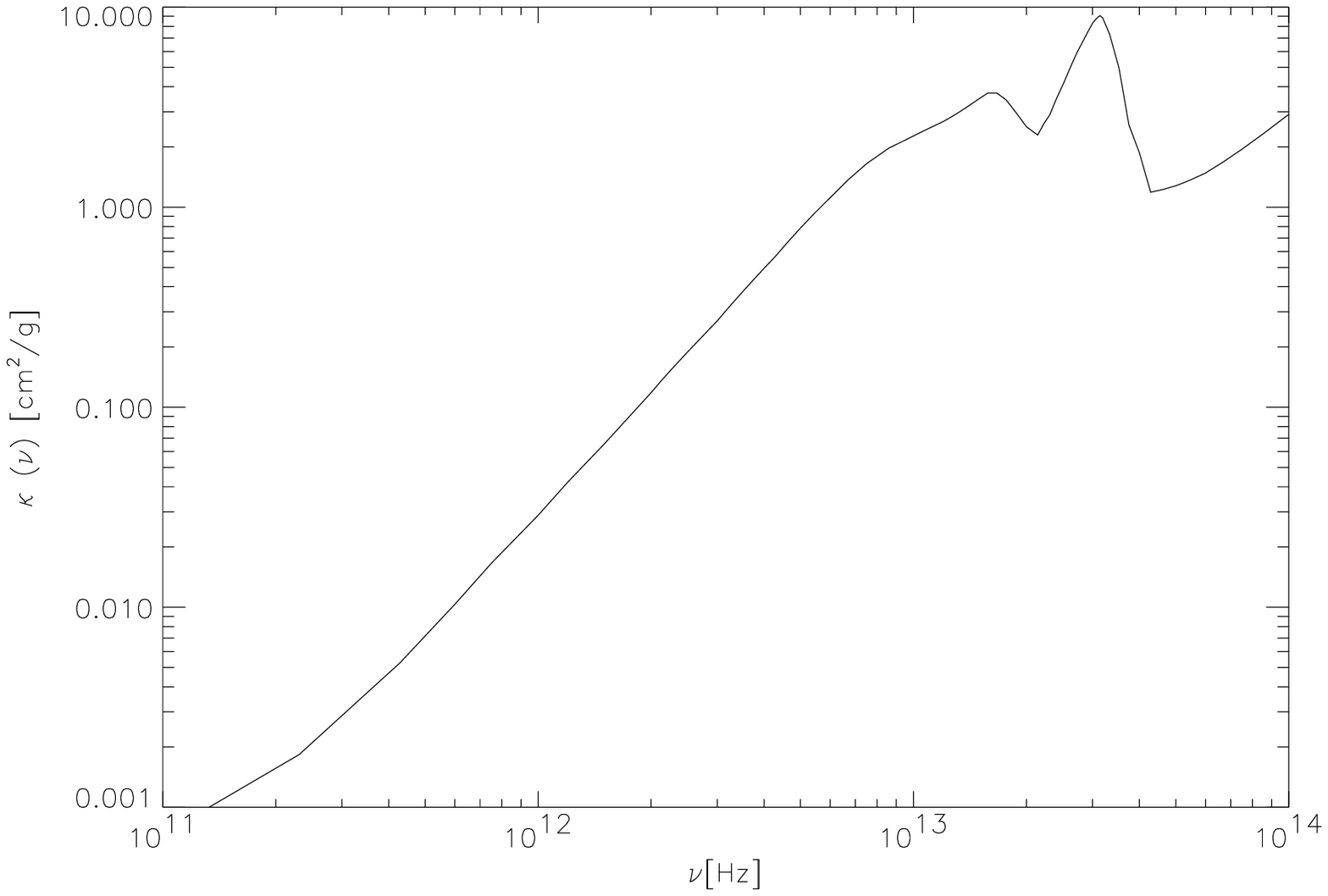,height=2.8in,width=2.8in}}}
\end{center}
\caption{(a) SED for typical high-mass protostar with frequency
regimes marked (solid line is DUSTY SED and crosses analytic SED). (b)
Contribution function (the characteristic emission radius) in
dimensionless units, $\tilde{r}_{m}$, with frequency regimes marked.
(c) WD01 opacity curve.  The spectral features in the SED and opacity
curve as shown in (a) and (c), e.g. the $3\times 10^{13}\rm Hz$ ($10
\micron$) absorption feature, correlate with the location in the
envelope this emission is coming from, as shown in (b)}
\end{figure}

\begin{figure}[!ht] \begin{center} \centerline{\psfig{file=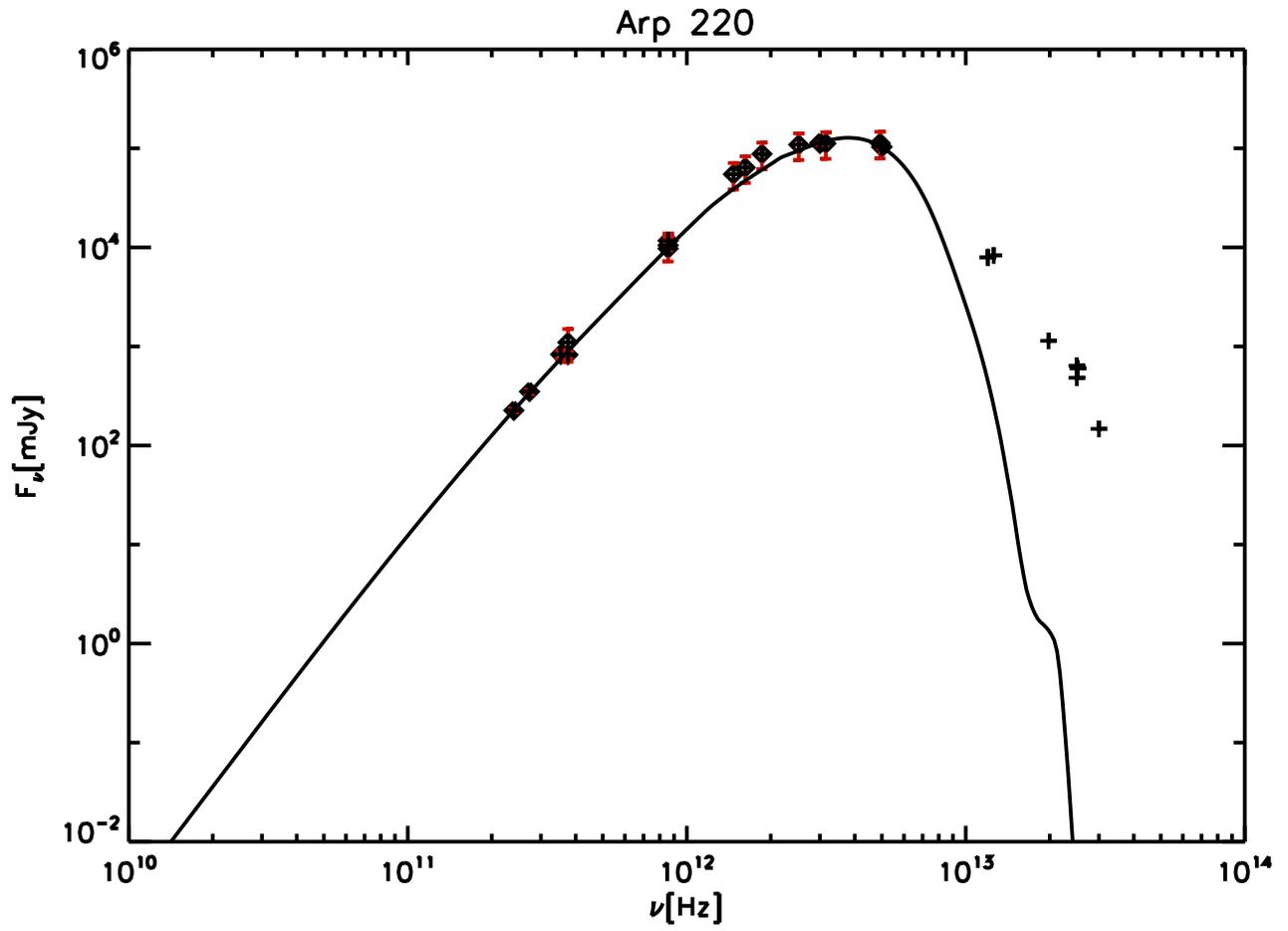}}
\end{center}
\caption{Best-fit SED of Arp 220}
\end{figure}

\begin{figure}[ht] \begin{center} \centerline{\psfig{file=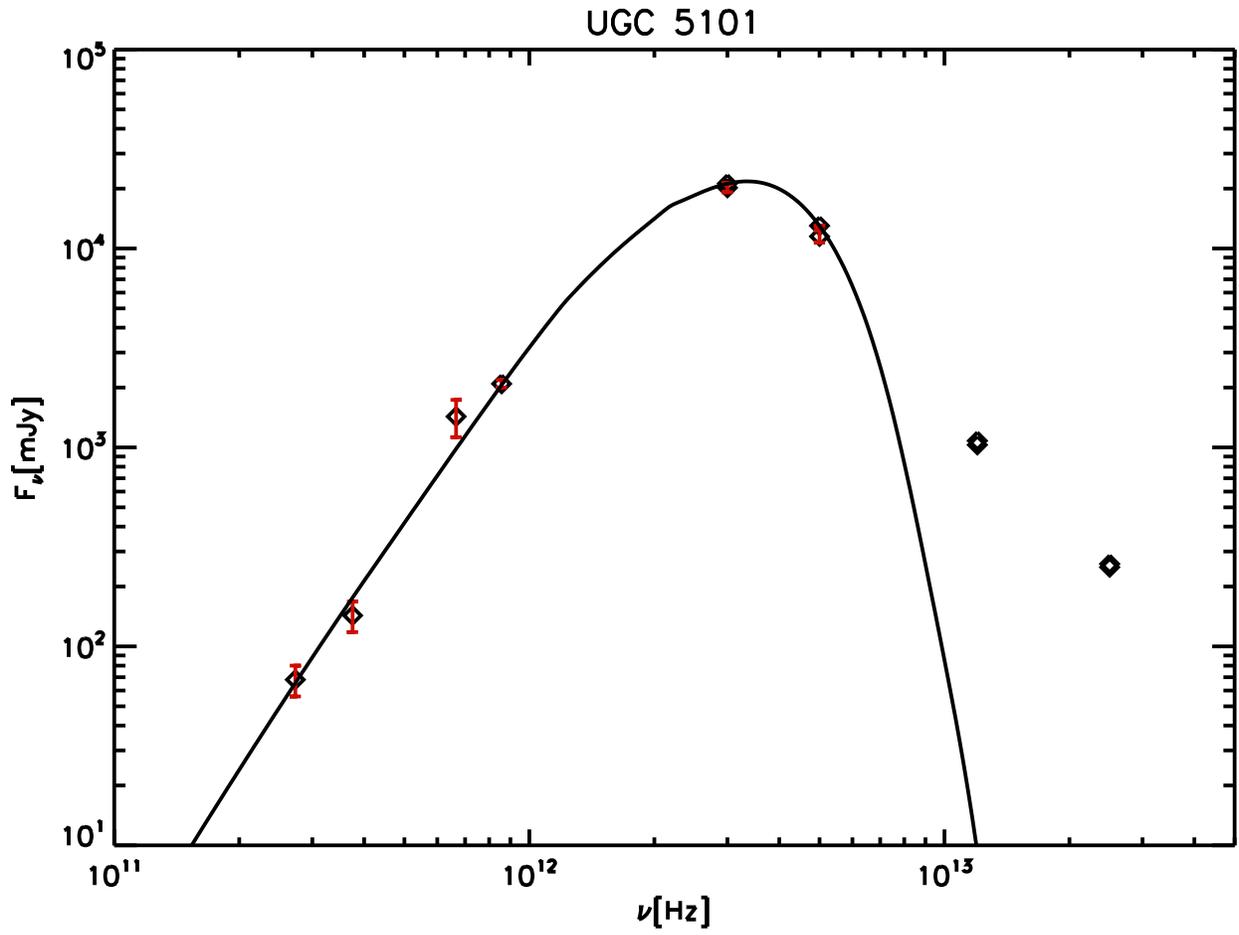}}
\end{center}
\caption{Best-fit SED of UGC 5101}
\end{figure}

\begin{figure}[ht] \begin{center}
\centerline{\psfig{file=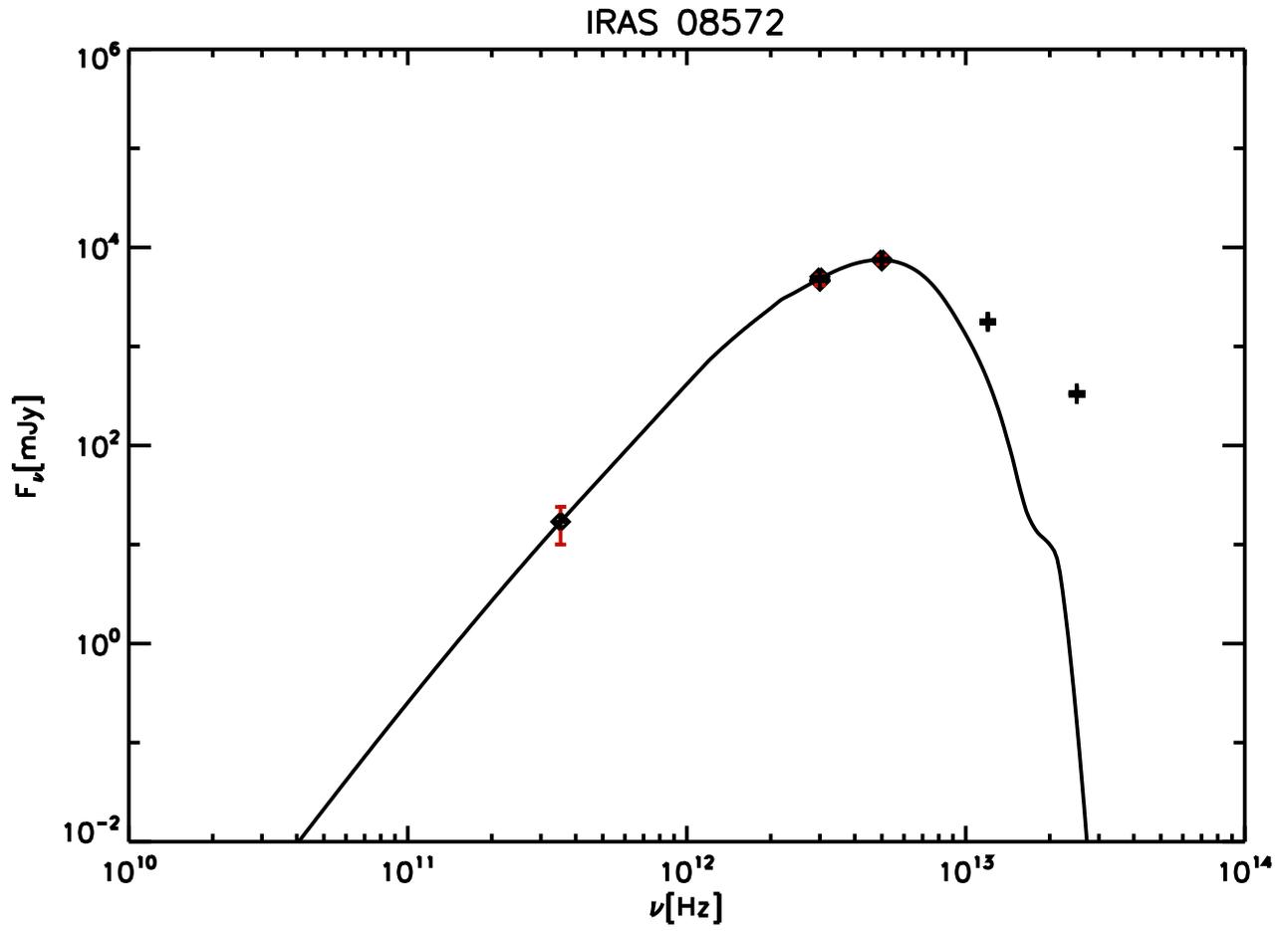}}
\end{center}
\caption{Best-fit SED of IRAS 08572+3915}
\end{figure}

\begin{figure}[ht] \begin{center} \centerline{\psfig{file=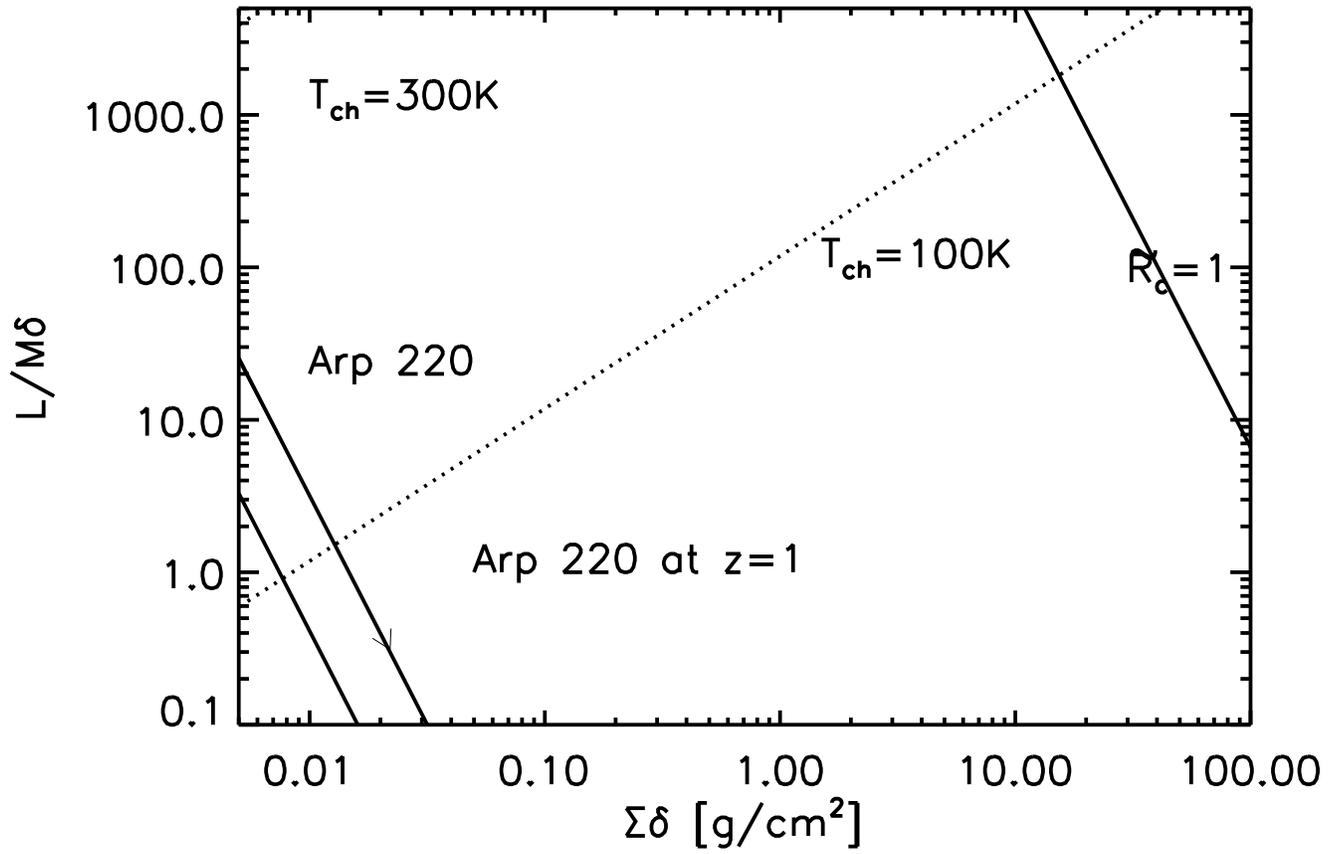}}
\end{center}
\caption{Relative change in inferred $L/M\delta$ and $\Sigma\delta$ as Arp 220 is
moved from $z=0.018$ to $z=1$ (arrow position is at z=1)}
\end{figure}

\begin{figure}[ht] \begin{center}
\centerline{\psfig{file=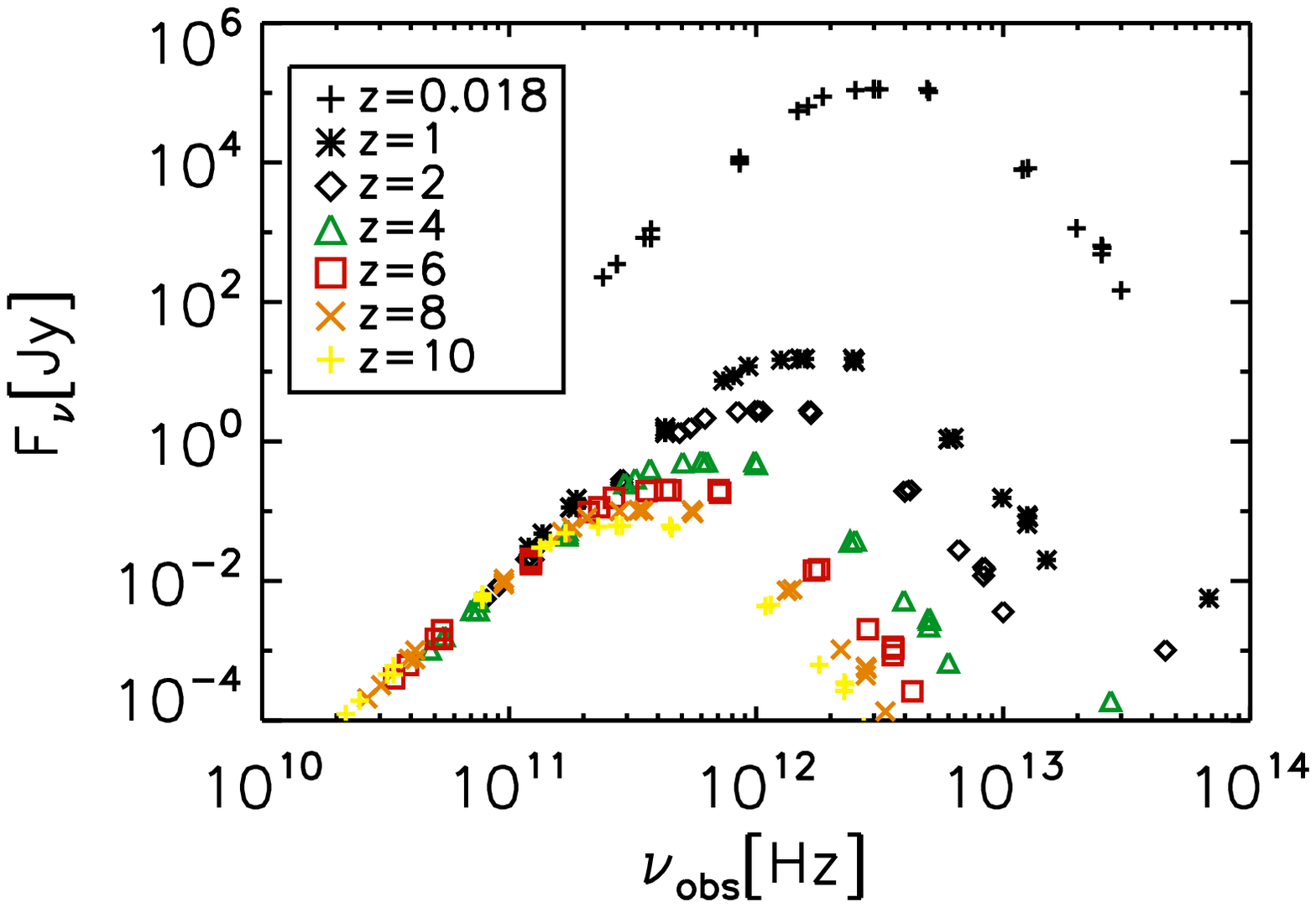,height=3.5in,width=3.5in}
\psfig{file=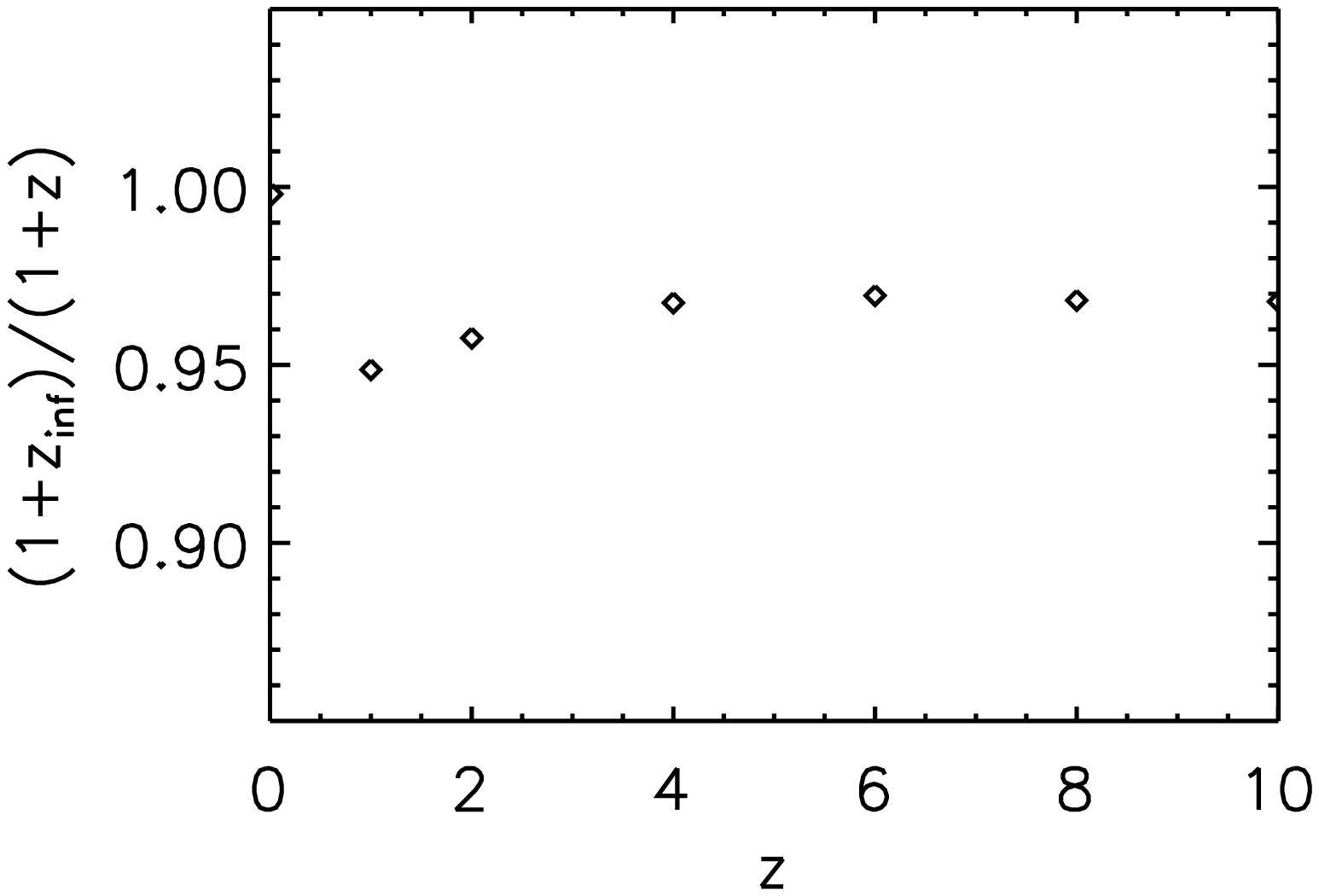,height=3.5in,width=3.5in}}
\end{center}
\caption{(a) SED of Arp220 shifted from $z=0.018$ to $z=10$ (b) Resultant inferred redshift compared to the actual redshift.}
\end{figure}

\clearpage

\begin{figure}[ht] \begin{center}
\centerline{\psfig{file=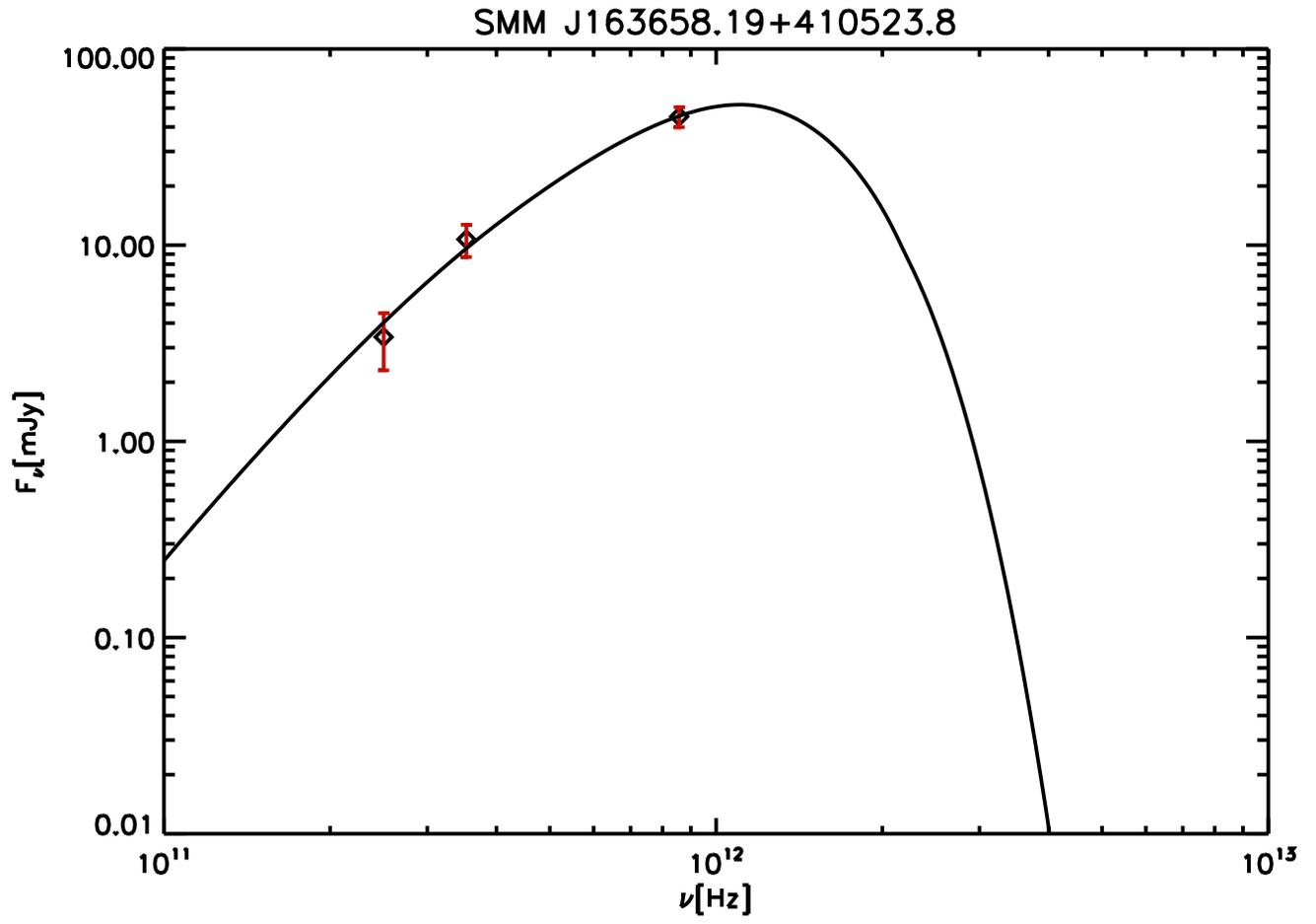}}
\end{center}
\caption{Best-fit SED of SMM J163658.19+410523.8}
\end{figure}

\begin{figure}[ht] \begin{center}
\centerline{\psfig{file=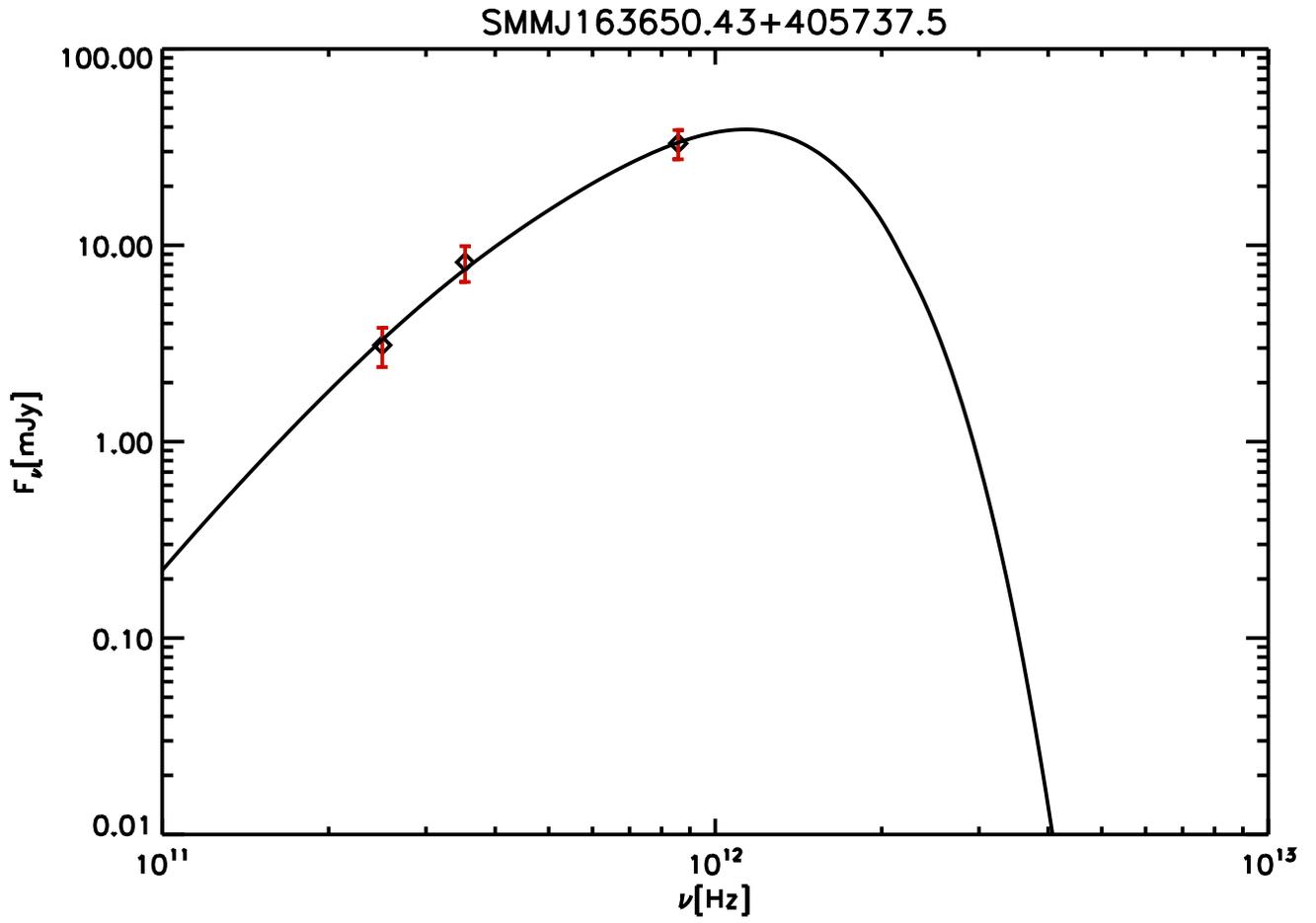}}
\end{center}
\caption{Best-fit SED of SMMJ163650.43+405737.5}
\end{figure}

\begin{figure}[ht] \begin{center}
\centerline{\psfig{file=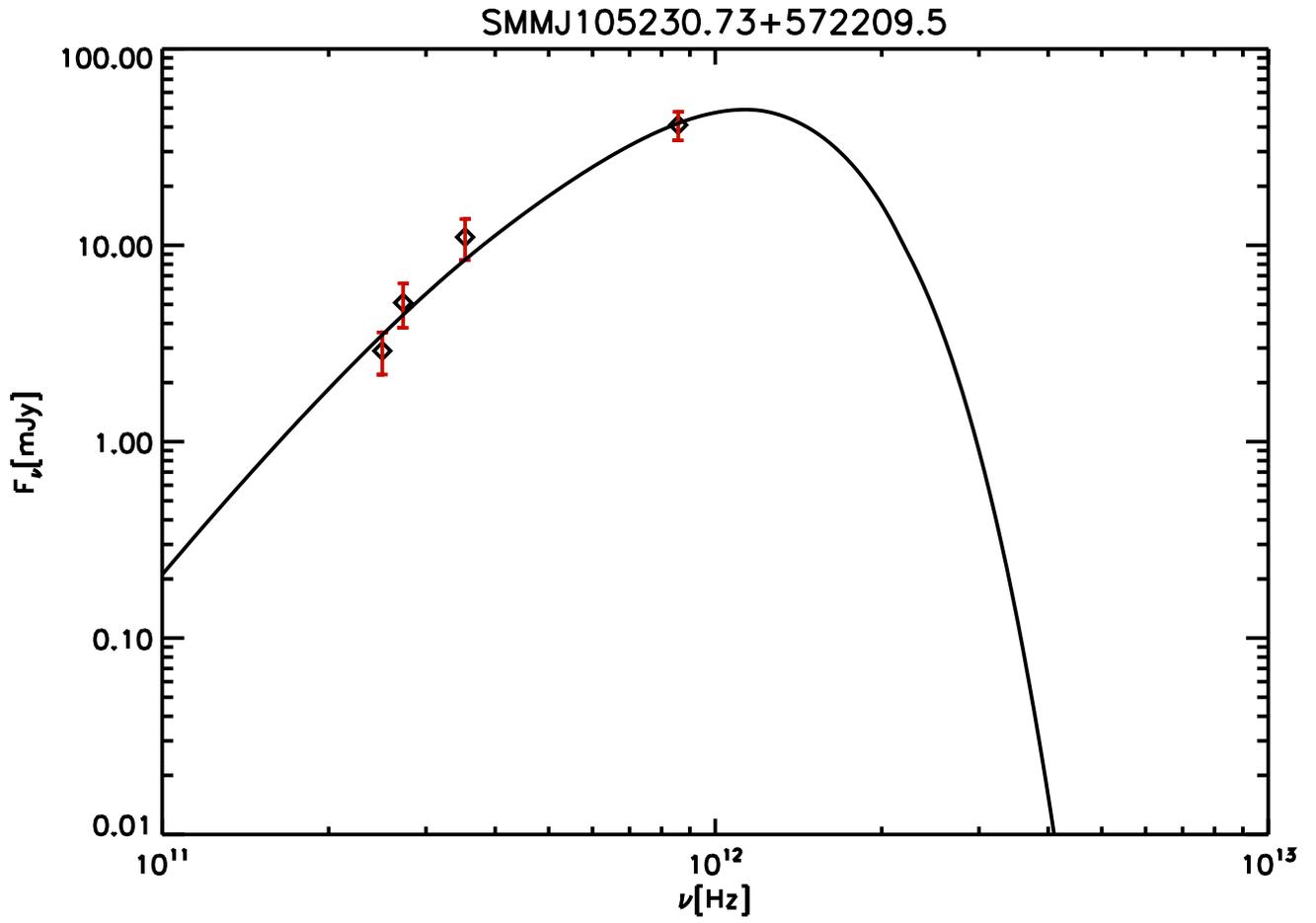}}
\end{center}
\caption{Best-fit SED of SMMJ105230.73+572209.5}
\end{figure}

\begin{figure}[ht] \begin{center}
\centerline{\psfig{file=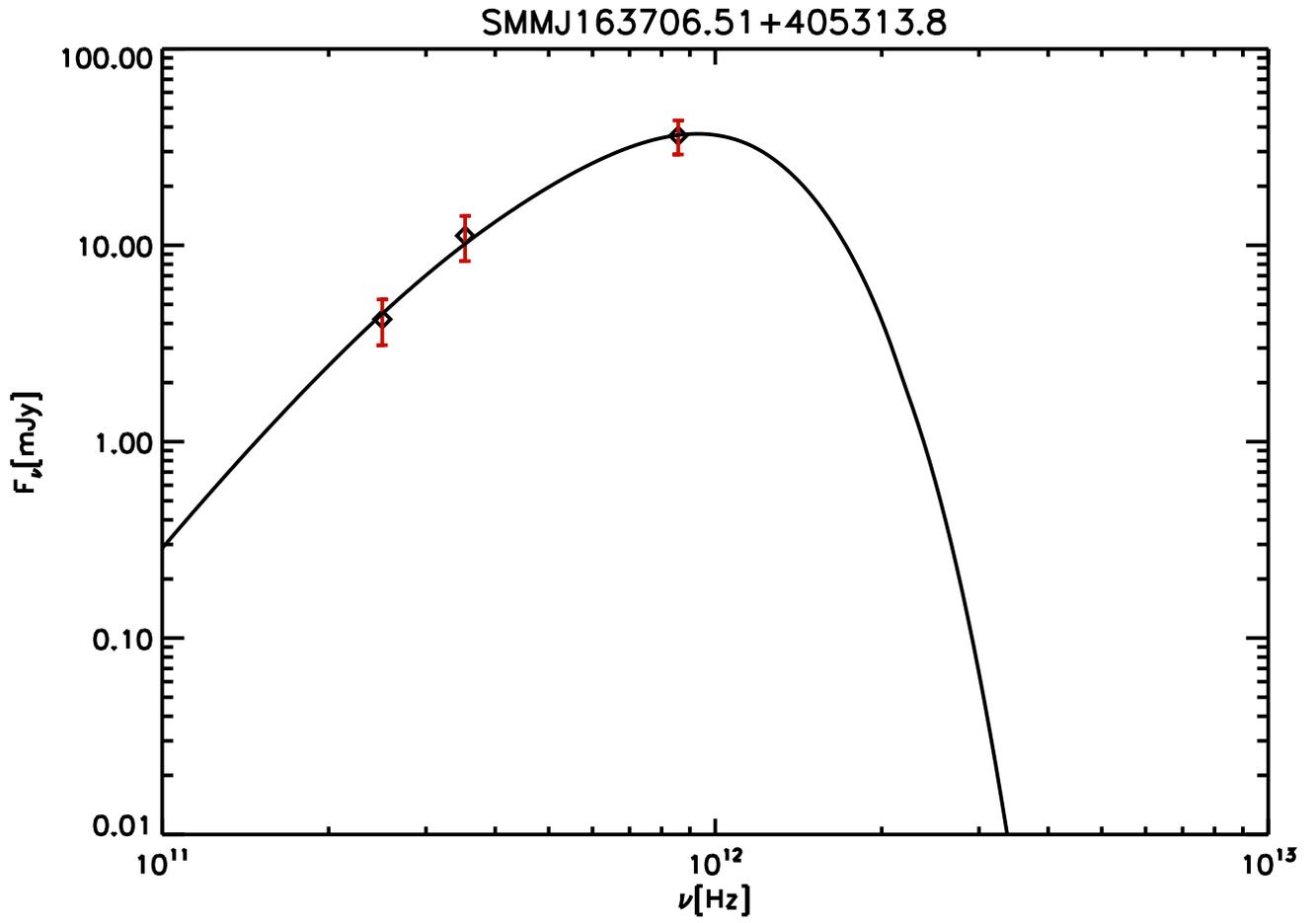}}
\end{center}
\caption{Best-fit SED of SMM J163706.51+405313.8}
\end{figure}

\begin{figure}[ht] \begin{center}
\centerline{\psfig{file=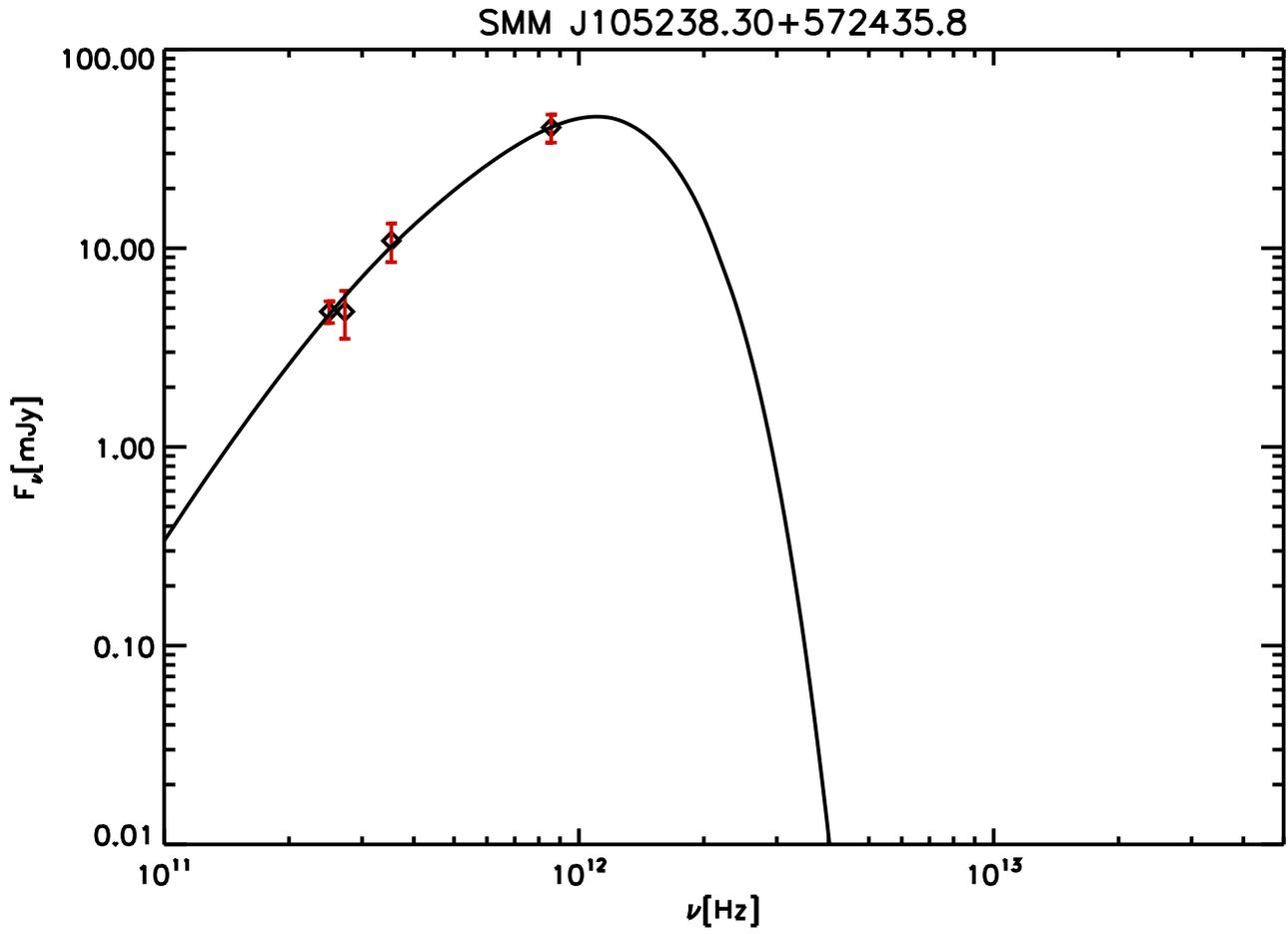}}
\end{center}
\caption{Best-fit SED of SMM J105238.30+572435.8}
\end{figure}


\begin{figure}[ht] \begin{center}
\centerline{\psfig{file=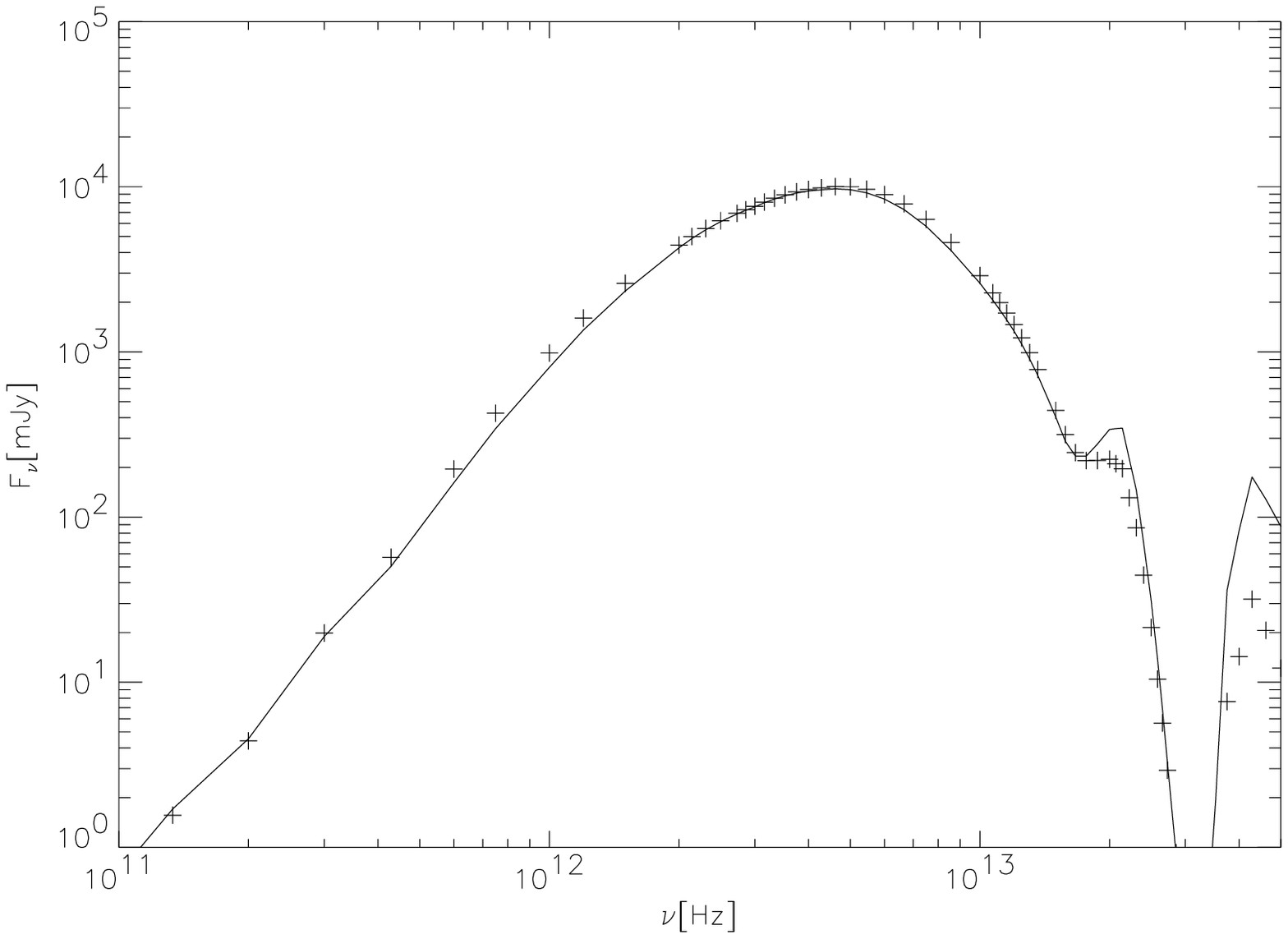,height=3.5in,width=3.5in}
\psfig{file=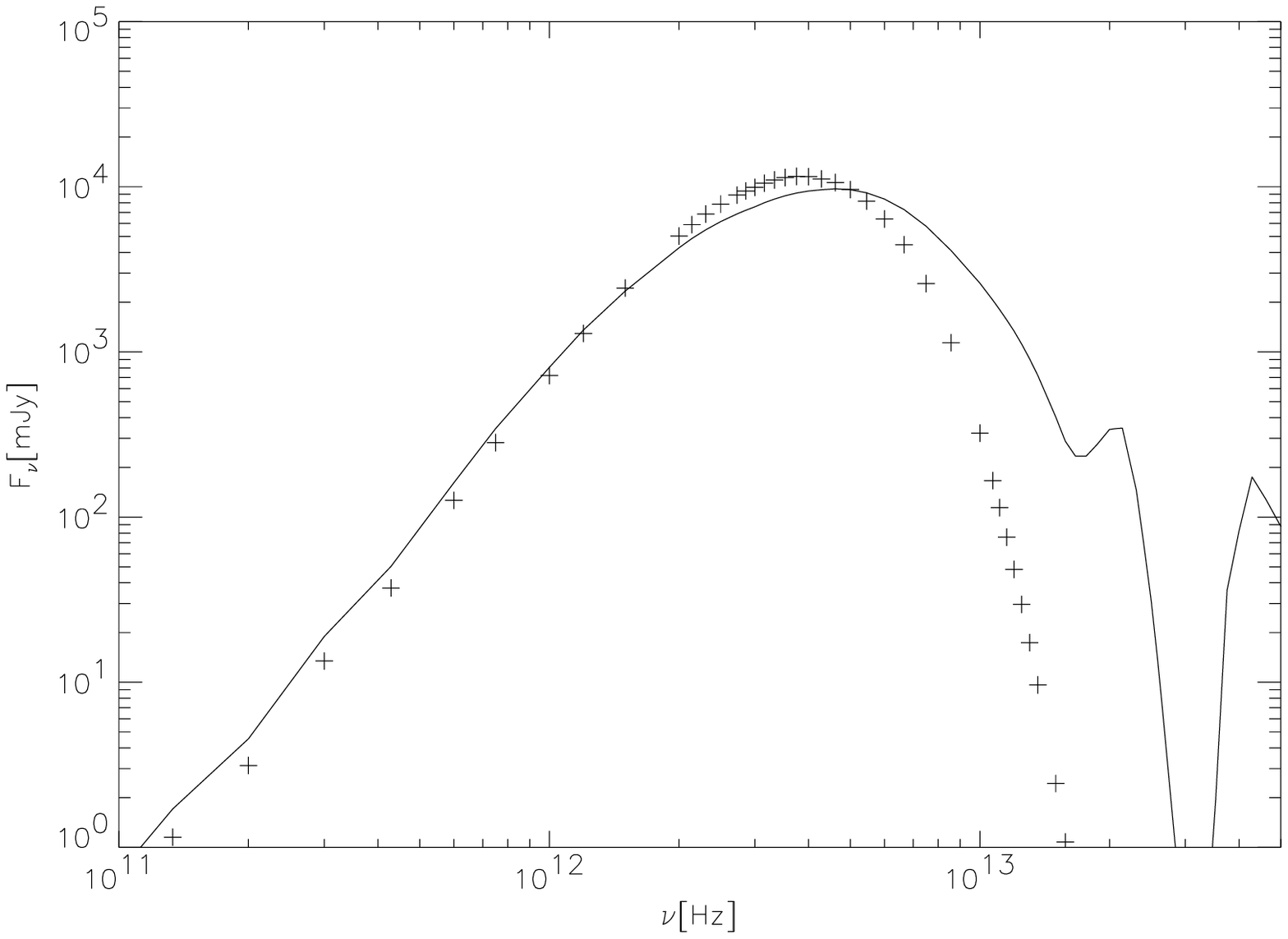,height=3.5in,width=3.5in}}
\end{center}
\caption{(a) SED for large $\rct\sim 300$; solid line is DUSTY, crosses
are analytic solution using methodology in CM05. (b) Solid line is
DUSTY, crosses are Hildebrand's prescription}
\end{figure}

\begin{figure}[ht] \begin{center}
\centerline{\psfig{file=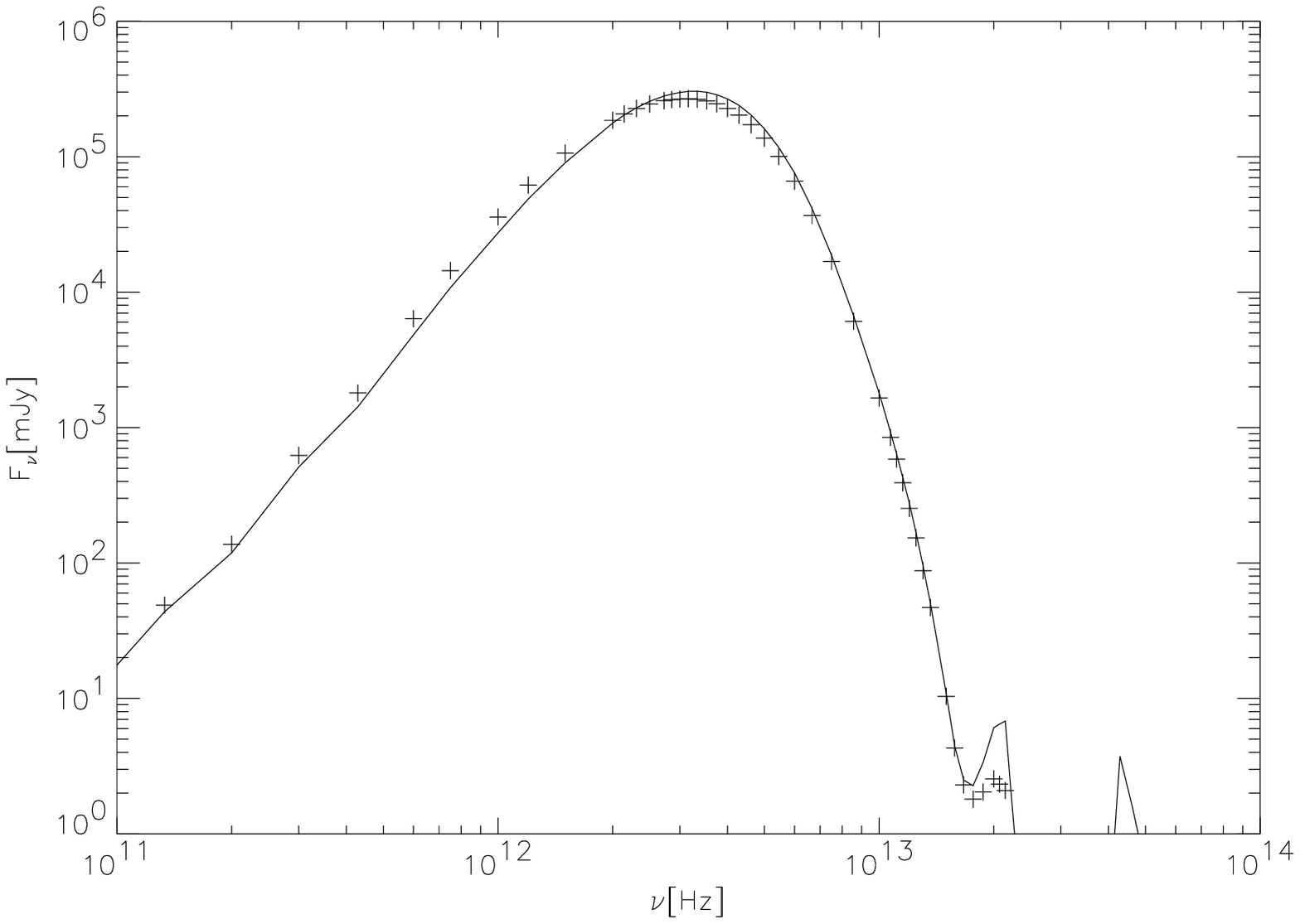,height=3.5in,width=3.5in}
\psfig{file=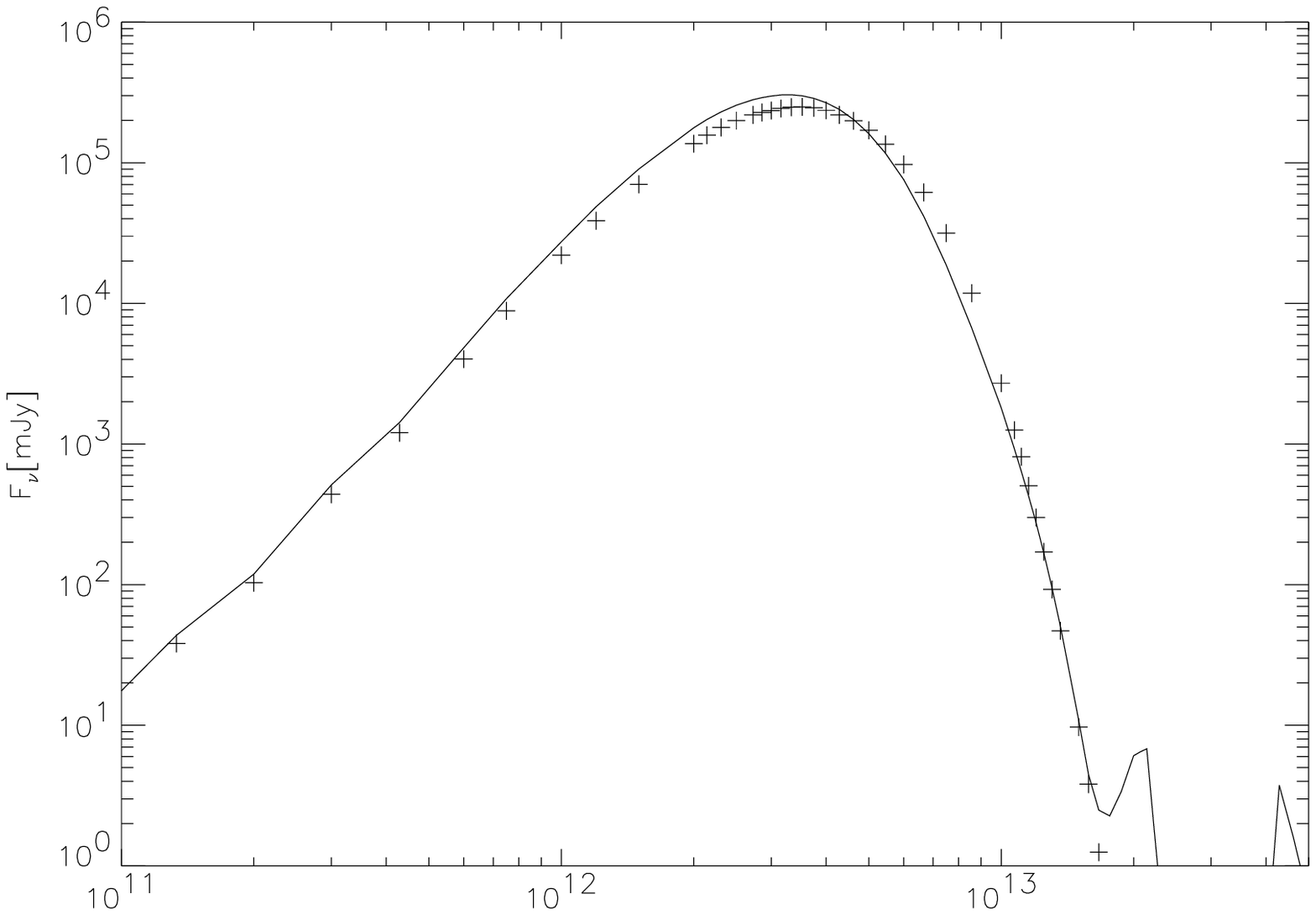,height=3.5in,width=3.5in}}
\end{center}
\caption{(a) SED for low $\rct\sim 10$; solid line is DUSTY, crosses are
analytic solution using methodology in CM05. (b) Solid line is DUSTY,
crosses are Hildebrand's prescription}
\end{figure}

\clearpage

\begin{deluxetable}{lcccccc} \tablewidth{0pt} \tablecolumns{7}
\tablecaption{Source Parameters of ULIRGs from SED}
\tablehead{\colhead{Source} & \colhead{$L\; (L_{\odot})$} & \colhead{$M\delta\;
(M_{\odot})$} & \colhead{$R_c$ (kpc)} & \colhead{$\Sigma\delta\; (\rm
g~cm^{-2})$} & \colhead{$\rct$} & \colhead{$\tch$}} 
\startdata 
Arp 220 & $9.4\times 10^{11} $ & $4.1\times 10^{10}$ & 10.6 & 0.025 & $370 \pm 20$ & $125 \pm 2.7$ \\ 
UGC 5101 & $7.1\times 10^{11} $ & $2.7\times 10^{10}$ & 5.5 & 0.07 & $164 \pm 24$ & $112 \pm 1.6$ \\ 
IRAS 08572+3915 & $8.9 \times 10^{11}$ & $6.9\times 10^{9}$ & 6.6 & 0.01 & $470 \pm 115$ & $172 \pm 4$ \\ 
Mrk 231 & $1.8 \times 10^{12}$ & $9.9 \times 10^{9}$ & 2.6 & 0.095 & $73 \pm 19$ & $138 \pm 4$ \\   
Mrk 273 & $9.7 \times 10^{11}$ & $4.1 \times 10^{10}$ & 13 & 0.016 & $509 \pm 19$ & $132 \pm 2.6$ \\
Mrk 1014 & $2.3 \times 10^{12}$ & $1.4 \times 10^{10}$ & 5.1 & 0.04 & $161 \pm 96$ & $153 \pm 4$ \\
IRAS 12112+3035 & $1.5 \times 10^{12}$ & $2.6 \times 10^{10}$ & 6.5 & 0.04 & $191 \pm 27$ & $131 \pm 2.6$ \\
IRAS 00262+4251 & $8.4 \times 10^{11}$ & $2.0 \times 10^{10}$ & 9.8 & 0.014 & $500 \pm 50$ & $145 \pm 9$ \\
IRAS 10565+2448 & $7.3 \times 10^{11}$ & $9.5 \times 10^{9}$ & 2.5 & 0.1 & $85 \pm 41$ & $122 \pm 2$ \\
IRAS 17208-0014 & $1.8 \times 10^{12}$ & $3.1 \times 10^{10}$ & 6.1 & 0.05 & $150 \pm 75$ & $128 \pm 2$ \\
\enddata
\end{deluxetable}

\begin{deluxetable}{lccccccc} \tablewidth{0pt} \tablecolumns{7}
\tablecaption{Source Parameters of SMGs from SED}
\tablehead{\colhead{Source} & \colhead{$L\; (L_{\odot})$} & \colhead{$M\delta\; (M_{\odot})$} & \colhead{$R_c$ (kpc)}  
& \colhead{$\Sigma\delta\; (\rm g~cm^{-2})$} &\colhead{$\rct$} & \colhead{$\tch$}}
\startdata 
SMMJ163658 & $6.8-8.2\times 10^{12}$ & $1.3-2.6\times 10^{11}$  & 5.3-16.0 & 0.3-0.07  & 40-150 & 103-115 \\ 
SMMJ163650 & $4.7\times 10^{12}$ & $1.85\times 10^{11}$ & 9.1 & 0.15 & $85 \pm 46$ & $102 \pm 7$ \\ 
SMMJ105230 & $7.2-9.3\times 10^{12}$ &  $1.1-2.3\times 10^{11}$ & 5.0-18.8 & 0.3-0.04 & 40-200 & 107-125 \\ 
SMMJ163706 & $4.3 \times 10^{12}$ & $1.6 \times 10^{11}$ & 3.8 & 0.74 & $25 \pm 16$ & $85 \pm 5$ \\
SMMJ105238 & $1.0 \times 10^{13}$ & $2.0\times 10^{11}$ & 13.4 & 0.074 & $122 \pm 65$ & $121 \pm 8$  \\
\enddata
\end{deluxetable}

\begin{deluxetable}{lccc} 
\tablewidth{0pt}
\tablecolumns{4}
\tablecaption{Inferred Redshifts for SMGs}
\tablehead{\colhead{Source} & \colhead{$(1+z_{\rm inf})/(1+z)$(ULIRG norm)} & \colhead{$(1+z_{\rm inf})/(1+z)$(SMG norm)}  & \colhead{$z_{\rm spec}$}} 
\startdata
SMMJ163658  & 1.02-1.1  &  0.96-1.04  & 2.454 \\
SMMJ163650  & 1.15      & 1.08        & 2.376 \\
SMMJ105230  & 0.98-1.06 & 0.92-1.00  & 2.611  \\
SMMJ163706  & 1.14      & 1.07        & 2.374  \\
SMMJ105238  & 1.03     & 0.96        & 3.036  \\
\enddata
\end{deluxetable}

\end{document}